\documentclass[a4paper,fleqn]{cas-sc}
\usepackage{framed}
\usepackage{tcolorbox}
\usepackage{multirow}
\usepackage{bibentry}
\usepackage{booktabs}
 \usepackage[numbers]{natbib}
 \usepackage[labeled]{multibib}
\newcites{PS}{Primary Studies}

\def\tsc#1{\csdef{#1}{\textsc{\lowercase{#1}}\xspace}}
\tsc{WGM}
\tsc{QE}
\tsc{EP}
\tsc{PMS}
\tsc{BEC}
\tsc{DE}

\begin{document}
\let\WriteBookmarks\relax
\def\floatpagepagefraction{1}
\def\textpagefraction{.001}

\shorttitle{Two Integration Pathways in Human-Centered Requirements Engineering}

\shortauthors{Benzarti et~al.}

\title[mode=title]{Two Integration Pathways in Human-Centered Requirements Engineering:
A Systematic Mapping Study of Structural Gaps}

\author[1]{Imen Benzarti}[
    orcid=0000-0003-0658-9605]
\cormark[1]
\ead{imen.benzarti@etsmtl.ca}
\credit{Conceptualization, Methodology, Investigation, Data curation, Writing -- Original draft}


\author[1]{Ikram Darif}
\ead{ikram.darif@etsmtl.ca}
\credit{Methodology, Investigation, Validation, Writing -- Review \& editing}


\author[2]{Abderrahmane Leshob}
\ead{leshob.abderrahmane@uqam.ca}
\credit{Validation, Writing -- Review \& editing}

\author[2]{Hafedh Mili}
\ead{mili.hafedh@uqam.ca}
\credit{Supervision, Writing -- Review \& editing}

\author[3]{Darine Amayed}
\ead{darine.amayed@uqac.ca}
\credit{Writing -- Review \& editing}

\affiliation[1]{organization={École de Technologie Supérieure},
    city={Montreal},
    state={Quebec},
    country={Canada}}

\affiliation[2]{organization={LATECE Laboratory, University of Quebec at Montreal},
    city={Montreal},
    state={Quebec},
    country={Canada}}

\affiliation[3]{organization={Department of Computer Science and Mathematics, University of Quebec at Chicoutimi},
    city={Chicoutimi},
    state={Quebec},
    country={Canada}}

\cortext[cor1]{imen.benzarti@etsmtl.ca (Imen Benzarti)}

\begin{abstract}
Human-centered Requirements Engineering (HC-RE) integrates user cognition, emotions,
and social interactions into the RE process through contributions from disciplines
such as psychology, cognitive science, design thinking, and human-computer interaction.
Despite growing interest, how these multidisciplinary contributions are structured and
why they remain fragmented across the RE lifecycle is not well understood.
 
This systematic mapping study analyzes 56 primary studies across seven
dimensions including RE phases, user involvement techniques, contributing disciplines,
and evaluation methods. Results show that 70\% of approaches involve multidisciplinary
contributions, yet only 39\% have been empirically evaluated and 48\% address only the
elicitation phase. A cross-study analysis reveals a structural separation between two
parallel integration traditions: a Cognitive-Formal (C-F) pathway grounded in
goal-based frameworks and formal modeling, and a Participatory-Iterative (P-I) pathway
grounded in scenario-based frameworks and iterative design. Each pathway has developed
complementary strengths, but their near-total disconnection explains the persistent
lifecycle concentration and theory-practice gap observed in the corpus.
 
The findings identify the absence of translation mechanisms between human-centered
artifacts and formal RE specifications as the field's primary structural gap, provide
a structured research agenda organized into four priority tiers, and establish the
empirical foundation for Experience-Centered Requirements Engineering, a direction
in which user experience is explicitly operationalized as a first-class concern in
requirements specification.
\end{abstract}

\begin{highlights}
\item Systematic mapping of 56 studies on human-centered approaches in RE
\item 70\% of approaches integrate multidisciplinary knowledge into RE
\item Four categories of discipline contributions to human-centered RE identified
\item Personas are the most widely used cross-discipline technique
\item Two-pathway taxonomy (Cognitive-Formal vs.\ Participatory-Iterative) 
      with zero framework crossover across 56 studies
\item Empirical foundation established for Experience-Centered 
      Requirements Engineering (XCRE)
\end{highlights}


\maketitle


\section{Introduction}

Requirements Engineering (RE) is a foundational phase of Software Engineering (SE),
concerned with defining and managing software requirements~\cite{8559686}. It focuses
on specifying \textit{what a system should do}, rather than \textit{how it should do it},
and plays a central role in ensuring that software systems align with stakeholder needs
and expectations. RE encompasses several interconnected activities, including
elicitation, analysis, specification, validation, and evolution of requirements
throughout the software lifecycle~\cite{van2000requirements,nuseibeh2000requirements}.
Its importance is widely recognized: deficiencies in RE often propagate to later
development stages, resulting in increased costs, project delays, and system
failures~\cite{chakraborty2012role}.

Beyond its technical dimension, RE is inherently a socio-technical activity in which
human factors play a critical role. Successful requirements engineering depends not
only on formal methods and technical accuracy, but also on the ability to understand
and integrate the perspectives, behaviors, and expectations of diverse
stakeholders~\cite{dalpiaz2013adaptive}. These stakeholders---including end users,
customers, domain experts, analysts, and developers---must collaborate to develop a
shared understanding of system goals~\cite{snijders2015crowd}. Consequently, the
consideration of human factors is not optional, but essential to effective RE.

In response to this need, recent research has increasingly emphasized
human-centered requirements engineering (HC-RE), which explicitly incorporates user
needs, emotions, and values into the RE process~\cite{boy2017human,kasauli2021requirements}.
A wide range of techniques has been proposed, including emotional goal modeling,
persona-based scenarios, co-creation workshops, and iterative user testing
~\cite{hussain2020human,sutcliffe2022implications,karolita2023use,boy2017human}.
These approaches are inherently multidisciplinary, drawing on fields such as
psychology, cognitive science, human-computer interaction (HCI), and design thinking.
As a result, they enable RE to account for cognitive load, emotional responses, and
social interactions alongside functional requirements
~\cite{leonardi2011design,boy2017human,sangiorgi2019human}.

Despite this diversity, a fundamental limitation remains. Existing secondary studies
tend to focus on isolated aspects of HC-RE, such as team-related human
factors~\cite{hidellaarachchi2021effects}, user involvement
patterns~\cite{abelein2015understanding}, or specific techniques such as
personas~\cite{wang2024uses}. However, there is still a lack of comprehensive
analysis of how multidisciplinary contributions are integrated across the RE
lifecycle. In particular, little is known about the structural reasons why
human-centered approaches are predominantly applied in early RE phases and remain
weakly connected to downstream activities such as specification and validation.
This limitation constrains the evolution of RE toward approaches in which user
experience is explicitly represented and operationalized as a first-class concern
in requirements specification. We refer to this direction as
\textit{Experience-Centered Requirements Engineering (XCRE)}.

To address this gap, this paper presents a systematic mapping study (SMS) of 56
primary studies. The objective is to analyze how multidisciplinary contributions are
structured within HC-RE, identify their limitations, and derive directions for more
integrated approaches. The main contributions of this study are as follows:

\begin{itemize}
    \item A systematic classification of 56 primary studies across seven analytical
    dimensions, including user involvement techniques, application domains, RE
    phases, frameworks, contributing disciplines, evaluation methods, and outcomes.

    \item A taxonomy of 11 contributing disciplines, organized into four categories
    based on their role in human-centered requirements engineering.

    \item A two-pathway taxonomy that reveals a structural separation between
    Cognitive-Formal and Participatory-Iterative integration traditions, with no
    crossover between framework types across the analyzed studies.

    \item A structured research agenda, organized into four priority tiers, that
    provides the empirical foundation for advancing Experience-Centered Requirements
    Engineering (XCRE).
\end{itemize}

The remainder of this paper is organized as follows.
Section~\ref{sec:background} presents the conceptual background on human-centered RE.
Section~\ref{sec:research-metho} describes the research methodology.
Section~\ref{sec:findings} reports the results of the mapping study.
Section~\ref{sec:gaps} identifies key research gaps.
Section~\ref{sec:recommendations} proposes a research agenda toward XCRE.
Section~\ref{sec:related-work} discusses related work.
Section~\ref{sec:threats-to-validity} examines threats to validity.
Finally, Section~\ref{sec:conclusion} concludes the paper.

\section{Background}\label{sec:background}

Requirements Engineering (RE) concerns the systematic identification, analysis,
specification, and validation of system requirements~\cite{nuseibeh2000requirements}.
Within this context, the notion of \textit{human factors} refers to the cognitive,
psychological, and social characteristics of individuals involved in or affected by
the RE process~\cite{grundy2021impact}. These factors include, among others,
emotions, motivation, cognitive load, communication styles, and personal values, all
of which influence how requirements are expressed, interpreted, and negotiated.

Research on human factors in RE has evolved along two complementary directions.
The first focuses on internal team dynamics, including personality traits,
motivation, communication patterns, and collaboration mechanisms among requirements
engineers and analysts~\cite{hidellaarachchi2021effects}. The second direction,
which is the focus of this study, examines how external disciplines contribute
methods and techniques that enhance the RE process by making it more responsive to
user needs, behaviors, and experiences. This perspective is commonly referred to as
human-centered requirements engineering (HC-RE). In this paper, we use the umbrella
term \textit{human-centered RE} to denote approaches that explicitly involve users
or address human factors—such as cognition, emotions, and usability—with the goal of
improving requirements elicitation, analysis, and validation.

A defining characteristic of HC-RE is its multidisciplinary nature. Disciplines such
as psychology and cognitive science provide theoretical models for understanding user
behavior and decision-making processes. Design thinking and user-centered design
(UCD) introduce iterative and participatory practices that structure user
involvement. Human-computer interaction (HCI) and user experience (UX) contribute
concrete artifacts, including personas, interaction models, and usability
guidelines, that help translate user insights into design-relevant representations.
More recently, emerging technologies such as virtual reality (VR) and large language
models (LLMs) have extended the range of techniques available for user involvement
and requirements elicitation. Understanding how these diverse contributions relate
to one another is essential for advancing HC-RE toward more coherent and effective
methodologies.

Despite this richness, the relationships between these disciplinary contributions
and their distribution across the RE lifecycle remain insufficiently understood.
In particular, it is unclear how these contributions interact, how they are
integrated across different RE phases, and why human-centered practices tend to
remain concentrated in early elicitation activities rather than being systematically
propagated into specification and validation. Addressing this limitation requires a
more structured understanding of how multidisciplinary approaches are organized and
applied in practice.

This study contributes to this understanding through a systematic analysis of 56
primary studies. The results reveal a structural separation between formal and
participatory integration traditions, which helps explain the limited lifecycle
integration of human-centered practices. This observation provides the empirical
foundation for \textit{Experience-Centered Requirements Engineering (XCRE)}, a
research direction in which user experience—including cognitive, emotional, and
social dimensions—is explicitly modeled and operationalized as a first-class concern
within requirements specification, rather than remaining implicit or confined to
early RE activities.

\section{Research Methodology}\label{sec:research-metho}

This Systematic Mapping Study (SMS) aims to identify, synthesize, and compare human-centered and user-centered approaches in Requirements Engineering (RE). The objective is to provide an in-depth analysis of existing research relevant to our research questions. To achieve this, we follow Petersen et al.'s guidelines for systematic mapping studies~\cite{petersen2015guidelines}, complemented by Kitchenham's general procedures for evidence
synthesis~\cite{kitchenham2004procedures}.

The SMS consists of three main phases: planning, conducting, and reporting.
During the \textbf{planning phase}, we define the SMS objectives and establish the review protocol. This protocol includes:  
(1) defining the research questions,  
(2) formulating the search query, and  
(3) identifying inclusion and exclusion criteria.  
In the \textbf{conducting phase}, we apply the review protocol to select relevant primary studies. The steps in this phase include:  
(1) conducting a search using the predefined search query,  
(2) selecting relevant primary studies based on predefined criteria, and  
(3) extracting and synthesizing data to answer the research questions.  
Finally, the \textbf{reporting phase} presents the SMS findings, highlights research gaps, and provides recommendations. The following sections detail the steps taken during the planning and conducting phases.

\subsection{Research Questions}

We developed the research questions using the PICOC framework (Population, Intervention, Comparison, Outcome, and Context), which is widely applied in empirical software engineering studies~\cite{wohlin2012experimentation, keele2007guidelines}. Based on the framework presented in Table~\ref{tab:PICOC}, we developed the following questions:

\begin{table}[]
\centering
\begin{tabular}{p{3cm}p{11cm}}
\hline
\textbf{Intervention} & Human-centered approaches in RE \\ \hline
\textbf{Outcomes}     & A structured understanding of human-centered RE approaches, including their impact on the RE process and their classification into distinct methodological categories. \\ \hline
\textbf{Context}      & Requirements Engineering \\ \hline
\textbf{Population} & Software engineering researchers and practitioners
applying or proposing human-centered approaches in RE \\ \hline
\textbf{Comparison} & Not applicable (mapping study; no controlled comparison  between specific interventions is performed)\\ \hline
\end{tabular}
\caption{PICOC framework for research questions.}
\label{tab:PICOC}
\end{table}
\begin{itemize}

 \item\textbf{RQ1:} \textit{Under what circumstances are human-centered approaches used?}  
This question examines the contexts in which human-centered approaches are applied in the RE process across different primary studies. It aims to identify techniques used to involve users in human-centered RE (e.g., interviews, surveys), application domains covered in primary studies (PSs), RE phases addressed in PSs (e.g., elicitation, specification), and RE frameworks employed (goal-based vs. scenario-based).  

 \item\textbf{RQ2:} \textit{Which disciplines contribute to human-centered RE?}  
This question investigates the disciplines that influence human-centered RE, including fields beyond computer science (e.g., psychology, cognitive science, social sciences) and sub-disciplines within computer science (e.g., Human-Computer Interaction (HCI), Virtual Reality (VR)).  

 \item\textbf{RQ3:} \textit{How do contributing disciplines complement the RE process?}  
This question explores the impact of different disciplines on RE, focusing on techniques they introduce to the RE process, their influence on RE practices, and how they contribute to transforming RE into a human-centered process.  

 \item\textbf{RQ4:} \textit{Which RE methodologies or techniques are used in human-centered approaches that do not rely on contributing disciplines?}  
This question identifies human-centered RE methodologies developed exclusively within RE, without the integration of external disciplines such as psychology, HCI, or UX. It focuses on methodologies that enhance user involvement, emotional considerations, and usability while remaining discipline-specific.  

 \item\textbf{RQ5:} \textit{How are human-centered RE approaches evaluated?}  
This question examines evaluation methods used to assess human-centered RE approaches and the outcomes derived from these evaluations.  
 
\end{itemize}
\subsection{Identifying Relevant Literature}

\subsubsection{Search Query}

To ensure a comprehensive selection of relevant primary studies, we incorporated alternative search terms (see Table~\ref{tab:alternative-search-terms}). These alternatives were identified through a pilot study to enhance coverage.

\begin{table}[h]
\centering
\begin{tabular}{p{4cm}p{10cm}}
\hline
\textbf{Human-centered} & Human-centered / Human-oriented / People-oriented / User-centered / User-centric \\ \hline
\textbf{Human factors}  & Human aspects \\ \hline
\textbf{Requirements Engineering} & Requirements Elicitation / Requirements Specification / Requirements Analysis \\ \hline
\end{tabular}
\caption{Alternative search terms.}
\label{tab:alternative-search-terms}
\end{table}

Using these terms, we formulated the following search query to focus on software engineering-related papers:

\begin{framed}
\textbf{Search Query:} \\
((( "human centered") OR ("human-centered") OR ("human factors") OR ("human oriented") OR ("people oriented") OR ("human aspects") OR ("user oriented") OR ("user centered") OR ("user-centric")) \\
AND \\
(("requirements engineering") OR ("requirements elicitation") OR ("requirements specification") OR ("requirements analysis"))\\
AND \\
("software")
)
\end{framed}

\subsubsection{Primary and Secondary Search Process}

To conduct the search, we selected \textbf{Engineering Village}\footnote{https://www.engineeringvillage.com/search/quick.url}, which uses the Compendex and Inspec databases. 
We chose Engineering village because it is widely used by the software engineering community, and it aggregates multiple major bibliographic databases, including IEEE, ACM, Springer, Elsevier, and other major publishers in software engineering and related disciplines (e,g., Darif et al.~\cite{darif2026cnl}). 

No time range was specified to allow an analysis of how these approaches have evolved over time. The search process was conducted using both automatic and manual methods: (1)~Automatic search: conducted using the Engineering Village search engine, (2)~Manual search: performed by scanning primary studies identified through the automatic search, and (3)~Backward and forward snowballing: used to identify additional relevant studies. The snowballing process followed Wohlin's guidelines~\cite{wohlin2014snowballing}. Both backward snowballing (examining reference lists of selected studies) and forward snowballing (identifying studies citing the selected papers) were performed. This process yielded 19 candidate studies, of which 9 met the inclusion criteria after full-text review. To mitigate the risk of missing relevant studies, we complemented the automatic search with this snowballing process, which contributed 9 additional papers not captured by the primary search, resulting in a final dataset of 56 primary studies.

\subsection{Paper Selection Criteria}

\subsubsection{Inclusion and Exclusion Criteria}

The selection of studies was conducted based on 4 inclusion criteria (see table~\ref{tab:inclusion-criteria}) and 5 exclusion criteria (see table~\ref{tab:exclusion-criteria}) to filter the papers found and ensure that the final articles were aligned with our review objectives and research questions. Specifically, we focused on studies written in English, more than 4 pages long, and which address human-centered RE from a user-focused perspective.  Zotero Library was used to maintain the relevant records of the papers from the initial step to the final screening. Since all included papers are peer-reviewed, we consider the peer-review process itself as a baseline quality control mechanism.

\begin{table}[]
\begin{tabular}{p{1cm}p{13cm}}
\hline
\textbf{ID} & \textbf{Criterion}                                                                                \\\hline
I1 & The paper's full text published as journal or conference paper that comply with RE.                          \\\hline
I2 & The paper is written in English.                                                          \\\hline
I3 & The paper proposes an approach for human centered RE.                                     \\\hline
I4 & The paper proposes an approach for human centered software engineering including RE phase.\\\hline
\end{tabular}
\caption{Inclusion criteria. \label{tab:inclusion-criteria}}
\end{table}

\begin{table}[]
\begin{tabular}{p{1cm}p{13cm}}
\hline
\textbf{ID} & \textbf{Criterion}                                                                                \\\hline
E1 & Less that 4 pages length short paper.                           \\\hline
E2 & A paper about RE but not discussing  human centered issues.                                                         \\\hline
E3 & A paper about human centered issues within development teams.                                   \\\hline
E4 & A paper about human centered software engineering not concerned with RE phase. \\\hline
E5 & Extended journal article of a same paper. \\\hline
\end{tabular}
\caption{Exclusion criteria. \label{tab:exclusion-criteria}}
\end{table}

\subsubsection{Filtering of the Papers}

The filtering process involved three screening stages. Initially, $551$ potentially relevant papers were retrieved from Engineering Village using the defined search string. After removing duplicates, $431$ unique papers remained, which were then assessed based on the inclusion and exclusion criteria. In the first screening, a selection based on titles and abstracts reduced the pool to $127$ papers for further analysis. The second screening involved a more detailed review of the introduction, methodology, results, and conclusions, further narrowing the selection to $67$ papers. Finally, in the third screening, additional refinement during the data extraction process led to the final inclusion of 47 papers.

After the filtering process, we obtained $47$ papers from our primary database search and $19$ papers from snowballing search, filtered to $9$ papers.  This resulted in a total of 56 included papers. We used the tool Rayyan~\footnote{https://www.rayyan.ai/} during the filtering process. The first and second authors performed the filtering independently in blind mode  and then compared their selections of the relevant papers.

\begin{table}[]

\begin{tabular}{lllll}
\hline
                 & Initial Paper Count & First Screening & Second Screening & Third Screening \\ \hline
Primary Search   & 551                 & 127              & 67               & 47              \\ \hline
Secondary Search & 19                  & 19              & 9                & 9               \\ \hline
\multicolumn{4}{l}{Total final paper count (Primary + Secondary)}           & 56              \\ \hline
\end{tabular}
\caption{Detailed Paper Count Analysis}
\label{tab:paper-count}
\end{table}

\subsection{Data Extraction Strategy}

To ensure consistency in data extraction, we developed a Google Form with 27 sections and 52 questions, listed in Appendix~\ref{annex:data-extraction}, covering key aspects of human-centered Requirements Engineering (RE). The form collected information on general details (title, authors, venue), research focus (goal, research questions, application domain), user and expert involvement, RE phases targeted, influence of contributing disciplines (e.g., HCI, UX, psychology), methodologies and techniques used, evaluation methods, and study outcomes.

A pilot study was conducted to refine the extraction process. The first and second authors independently extracted data from 10 randomly selected papers in blind mode using the standardized form. Discrepancies were discussed until consensus was reached. After three iterations of refinement, the final version of the extraction form was validated.

The classification categories used in the analysis (e.g., contributing disciplines, RE phases, application domains, technique categories) were developed inductively through iterative open coding. During the extraction process, both authors independently assigned descriptive labels to each paper based on its content. These labels were then consolidated through discussion, grouping related codes into higher-level categories. For example, the four categories of discipline contribution in RQ3 --- Understanding Human Cognition, Enhancing User-Centered Design, Improving User Satisfaction, and Simulating/Automating Elicitation --- emerged from clustering the techniques and contributions observed across the 39 discipline-influenced studies. Similarly, the three categories of RE-internal approaches in RQ4 (User-Centric Requirements, Social and Organizational Contextualization, and Linguistic and Cognitive Workload Considerations) were derived by grouping the techniques proposed by the 17 studies that did not rely on external disciplines. This inductive approach follows established practices for qualitative data analysis in systematic mapping studies~\cite{petersen2015guidelines}.

The first and second authors conducted the screening independently in blind mode using the Rayyan tool\footnote{https://www.rayyan.ai/}. After independent screening, inclusion decisions were compared and disagreements were resolved through discussion. In cases where consensus could not be reached, a third author was consulted for final adjudication. To support transparency and reproducibility, we provide a replication package publicly available at~\cite{benzarti2026replication}. The package includes the complete list of retrieved papers with inclusion and exclusion decisions at each screening stage, the data extraction form, and the completed extraction data for all 56 primary studies.



\section{Findings}\label{sec:findings}
\subsection{Preliminary Findings}
 
\begin{figure}[]
  \centering
  \includegraphics[width=0.7\linewidth]{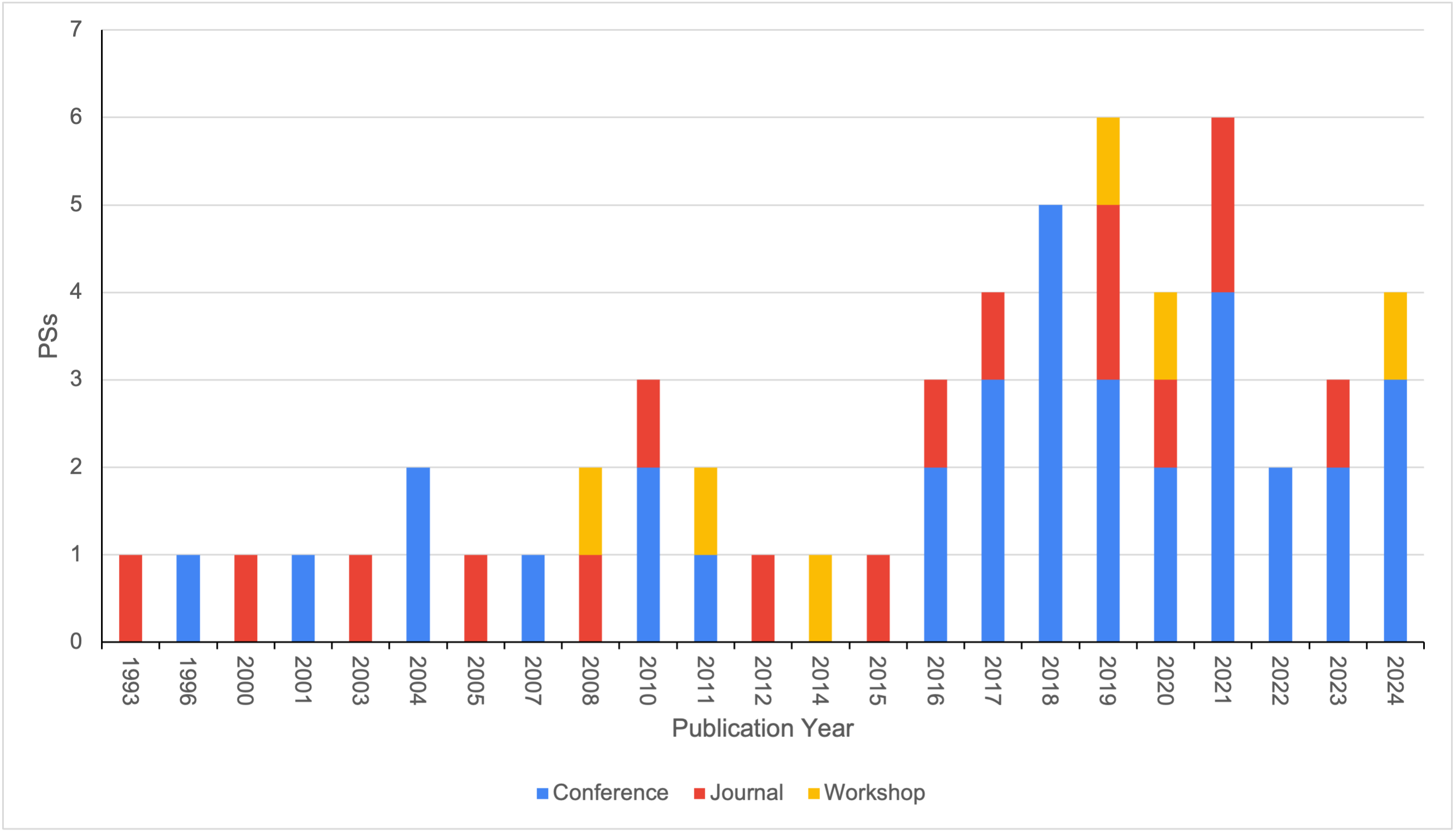}
  \caption{Distribution of PSs by year}
  \label{fig:publications_per_year}
\end{figure}

\begin{figure}[]
  \centering
  \includegraphics[width=0.5\linewidth]{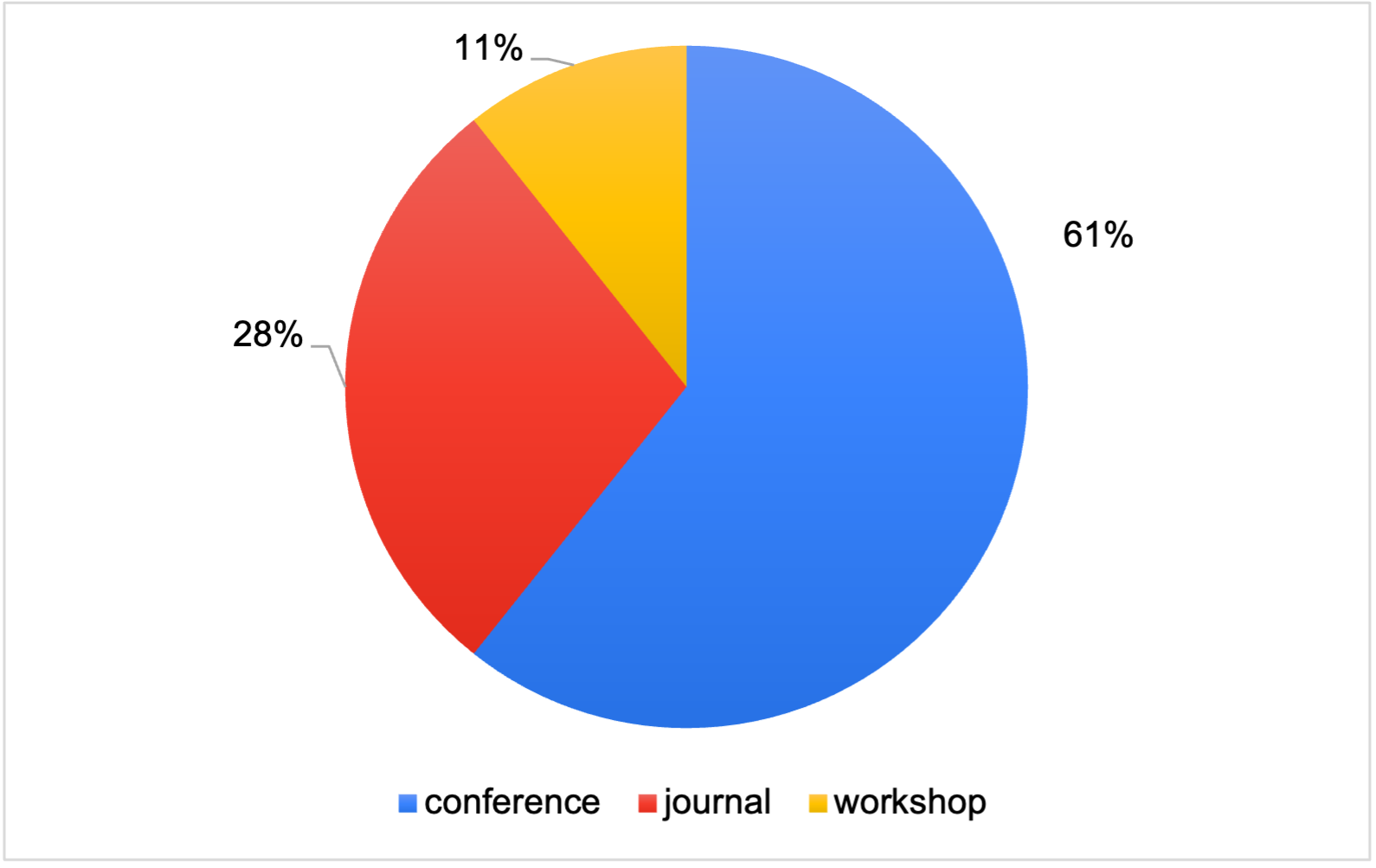}
  \caption{Distribution of PSs by study type}
  \label{fig:publications_per_type}
\end{figure}

Figure~\ref{fig:publications_per_year} shows publication trends over time. The first
primary study appeared in 1993, with sporadic contributions until 2010, after which
publications rose steadily. Studies from 2016--2024 account for 53\% of the corpus
(27 studies, including 7 journal papers), with peak years in 2019 and 2021 (six
publications each). As shown in Figure~\ref{fig:publications_per_type}, conferences
dominate as the primary publication venue, while journals and workshops contribute to
a lesser extent. The 56 PSs span 47 different venues covering requirements engineering,
software engineering, HCI, health informatics, design, and cognitive science ---
reflecting the inherently multidisciplinary nature of the research area and
corroborating the discipline taxonomy developed in RQ2. Regarding how authors
characterize their work, 39\% of PSs use the term user-centered (or user-centric),
22\% use human-centered, and 29\% carry no explicit qualification. Our analysis finds
no significant analytical distinction between these terms; they are used interchangeably
throughout this paper.
 
\subsection{RQ1: Under what circumstances are human-centered approaches applied?}
In this research question, we study the circumstances where the human-centered approaches are applied. We focus on user involvement techniques, the applications domains, covered RE phases, and used RE frameworks.

\subsubsection{How are users involved in the human-centered RE process?}
\begin{figure}[]
  \centering
  \includegraphics[width=0.7\linewidth]{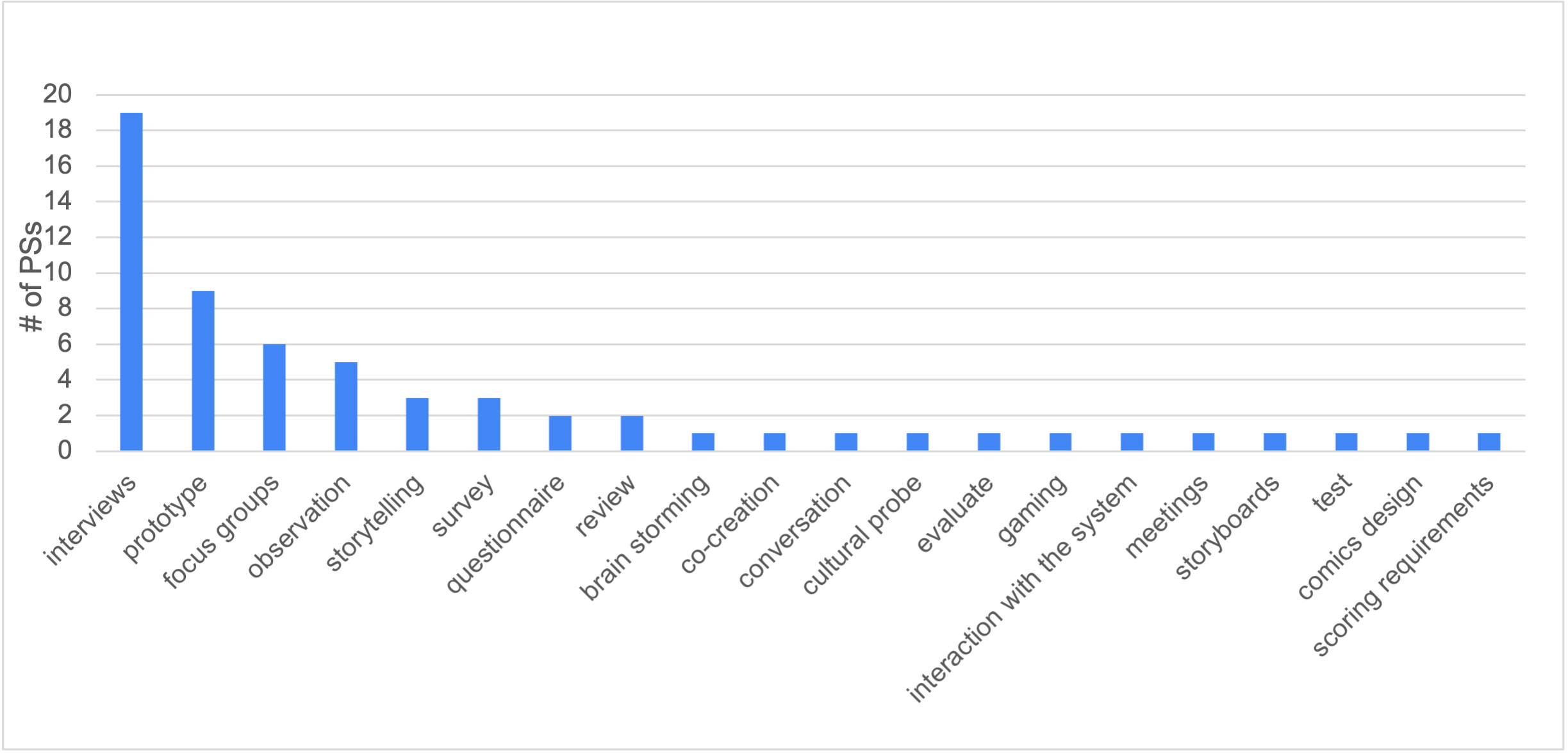}
  \caption{Number of PSs per Techniques used to involve users in RE process }
  \label{fig:PSper-user-involv}
\end{figure}

User involvement in the human-centered RE process ensures that the system accurately reflects their needs and expectations. Among the $56$ primary studies, users were involved in the RE process in 61\% (34 studies) 
Studies that involve users used a variety of research techniques. The most frequently used techniques were interviews (37.25\%), prototypes (17.64\%), focus groups (11.76\%), and observation (9.80\%). Less commonly used techniques included storytelling and surveys (5.88\% each), questionnaires and reviews (3.92\% each), with even fewer studies utilizing methods like gaming and co-creation (see Figure~\ref{fig:PSper-user-involv}). Among 30 studies, 14 used more than one technique to involve users and 6 used more than two techniques \citePS{PS4,PS12,PS18,PS25,PS31,PS39}. For instance, the approach in \citePS{PS25} combined prototype, interviews, focus groups, observation and conversation, and the approach in \citePS{PS39} combined prototype, storytelling, interviews, and storyboards. 

The dominance of interviews and prototypes reflects their maturity and ease of integration into existing RE practices. These techniques are well-established and require limited methodological adaptation. However, the limited use of more interactive approaches such as co-creation and gaming suggests that more participatory and immersive techniques remain underexplored. This may be due to higher implementation complexity, lack of methodological guidance, or limited tool support.


\subsubsection{What application domains are covered?}

\begin{figure}[]
  \centering
  \includegraphics[width=0.7\linewidth]{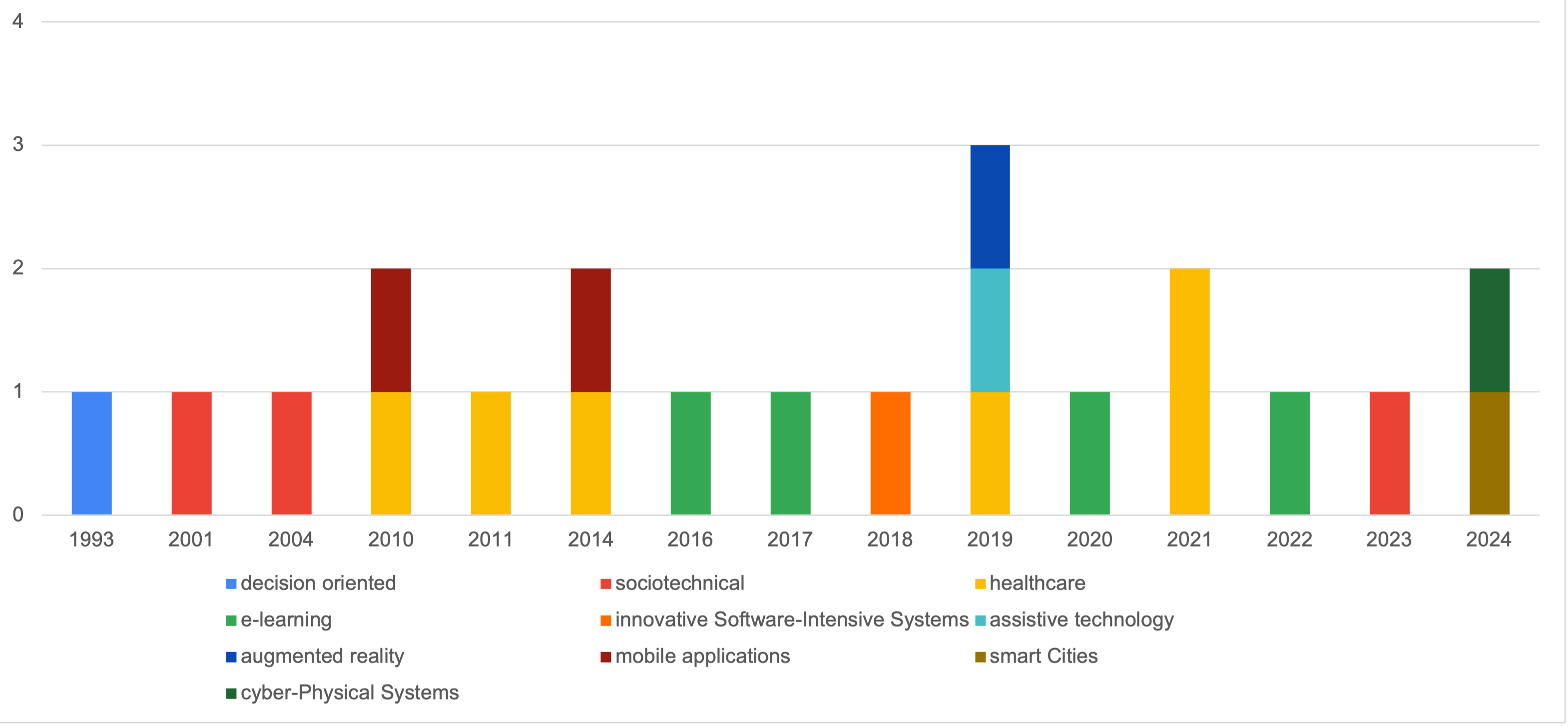}
  \caption{Number of PSs per application domain}
  \label{fig:Domains}
\end{figure}

Among the $56$ primary studies, 35 (62\%) are generic and 21 (36\%) are domain-specific. 
The most common application domains are healthcare (6 studies), e-learning (4 studies), and socio-technical systems (3 studies) (see Figure~\ref{fig:Domains}). These domains involve strong interactions between software and human actors and emphasize the importance of considering human factors in the design and development process. The prominence of these domains can be explained by their sensitivity to human interaction, where user behavior and experience directly impact system effectiveness. However, the limited representation of other domains suggests that the applicability of human-centered approaches across a broader range of contexts remains insufficiently explored. This imbalance indicates that current research may prioritize domains where human factors are already well recognized, potentially limiting the transfer of human-centered practices to less explored domains~\cite{storey2020software}.


\subsubsection{Which RE phases are covered?}

\begin{figure}[]
  \centering
  \includegraphics[width=0.5\linewidth]{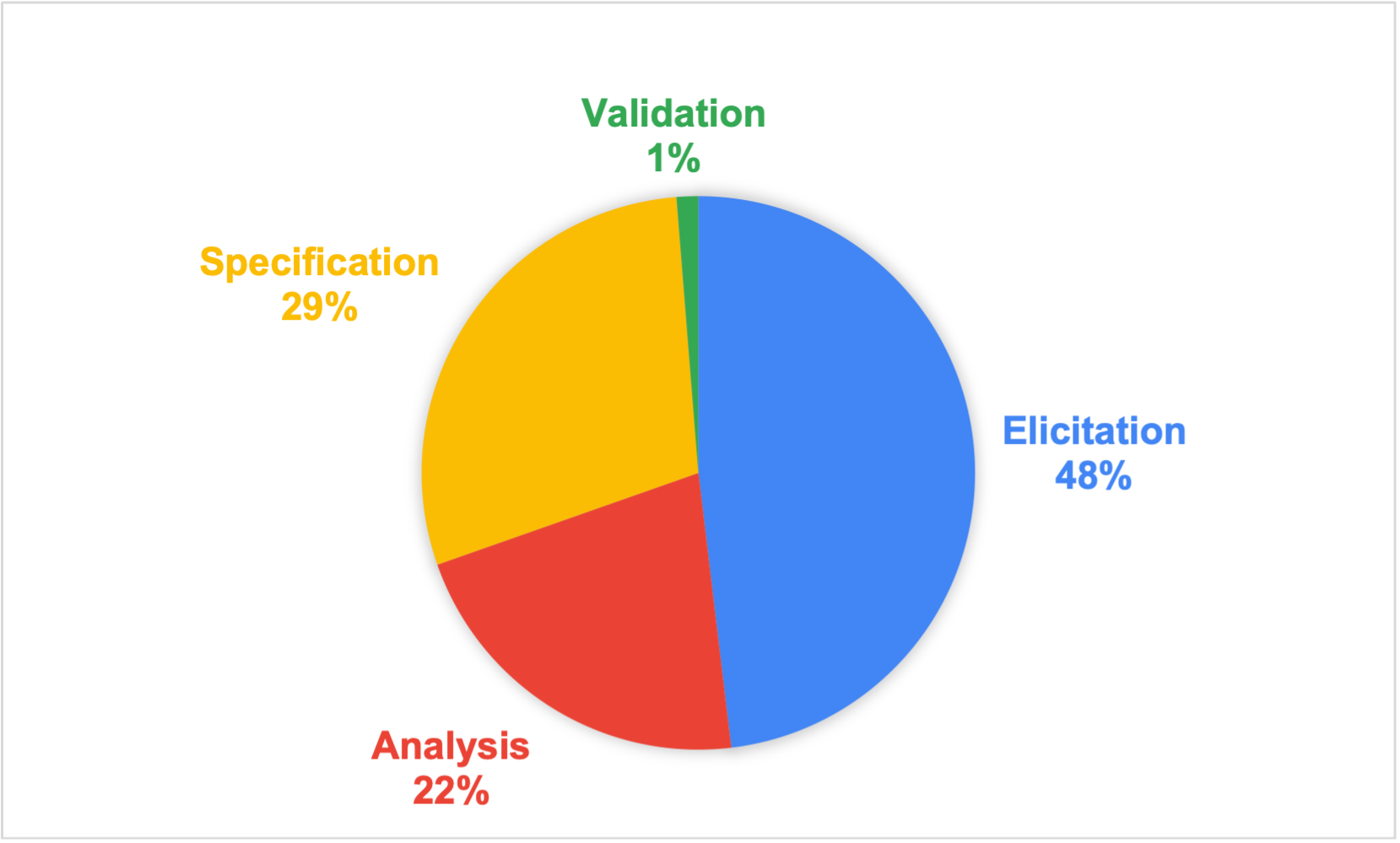}
  \caption{RE phases covered by the PSs}
  \label{fig:REPhase}
\end{figure}

Among the $56$ primary studies, 49 (88\%) were specific to the RE process, while 7 (12\%) covered the entire software engineering life cycle. As illustrated in Figure~\ref{fig:REPhase}, 48\% of the primary studies focus on the elicitation phase, which involves user participation to identify their requirements. Among the 49  RE-specific studies, 27 covered only one phase, while the remaining studies addressed two or more phases. 

The strong concentration on the elicitation phase suggests that human-centered approaches are primarily applied during early stages of RE, where direct interaction with users is most accessible. However, this also indicates a limited integration of human-centered practices in later phases such as specification, validation, and maintenance. This imbalance implies that while user needs are actively captured, they are not systematically preserved and operationalized throughout the full RE lifecycle.


Among the 56 primary studies (PSs), 20 used an existing requirements engineering (RE) framework. As shown in Table~\ref{tab:extended-RE-frameworks}, goal-based frameworks are the most utilized, with 11 PSs, while the remaining studies primarily rely on scenario-based frameworks. Most studies extend existing frameworks rather than proposing entirely new ones. Specifically, all PSs that utilized scenario-based frameworks introduced extensions, while five PSs extended goal-based models.

This preference for extending rather than redesigning existing frameworks reflects an
incremental integration strategy. More importantly, the framework type reveals a
structural alignment that anticipates the cross-pathway taxonomy in
Section~\ref{sec:cross-rq-synthesis}: without exception, every study using a
goal-based framework draws on formal and cognitive disciplines (psychology, cognitive
science, RE-internal emotional modeling), while every study using a scenario-based
framework draws on participatory disciplines (design thinking, UCD, HCI). This
zero-crossover alignment constitutes one of the clearest structural signals in the
corpus. However, regardless of framework type, most adaptations remain concentrated
in early RE phases, with limited propagation of human-centered concepts into
specification, validation, or maintenance.

\begin{table}[!h]
\begin{tabular}{ l l l l }
\hline
\textbf{Framework Type}         & \textbf{Framework}        & \textbf{Extended}     & \textbf{Not extended} \\ \hline
\multirow{6}{*}{Goal based}     & Contextual goal modelling & \citePS{PS3}                   &                       \\ \cline{2-4} 
                                & GRL                       & \citePS{PS26}                  &                       \\ \cline{2-4} 
                                & I*                        &                       & \citePS{PS36,PS42}            \\ \cline{2-4} 
                                & Agent-oriented goal model & \citePS{PS30,PS34}            &                       \\ \cline{2-4} 
                                & Motivational goal model   &                       & \citePS{PS23}, \citePS{PS33,PS45}      \\ \cline{2-4} 
                                & Tropos                    & \citePS{PS49}                  & \citePS{PS20}                  \\ \hline
\multirow{3}{*}{Scenario based} & Agile                     & \citePS{PS7,PS21,PS25,PS27} &                       \\ \cline{2-4} 
                                & Scrum                     & \citePS{PS1,PS4,PS9}         &                       \\ \cline{2-4} 
                                & Use cases                 & \citePS{PS41,PS47}            &                       \\ \hline
\end{tabular}
\caption{Extended RE frameworks in PSs}
\label{tab:extended-RE-frameworks}
\end{table}

\begin{tcolorbox}[colback=gray!10, colframe=black, title=Lessons Learned from RQ1]
Human-centered RE is predominantly applied as an elicitation activity, with limited integration across later RE phases. This reveals a lack of continuity in translating user insights into specification and validation, highlighting the need for end-to-end human-centered RE approaches.
\end{tcolorbox}

\subsection{RQ2: Which disciplines contribute to human-centered RE?}

RE is a multidisciplinary field that integrates principles from various disciplines to better understand user needs and interactions. This research question aims to explore the contributing disciplines that impact the human-centered RE process. 

Among $56$ primary studies (PSs), 39 PSs (70\%) are influenced by contributing disciplines. This high proportion indicates that human-centered RE rarely relies solely on traditional RE methods and instead depends on external knowledge domains to capture and model human aspects.
Within these 39 studies, the user-centered design emerges as the most popular field, with 10 studies incorporating its principles, as shown in Table~\ref{tab:texternal-descipline}. Psychology stands out as the second most common discipline, influencing 7 studies. Design thinking and human-computer interaction (HCI) contribute to 5 studies. The dominance of user-centered design and psychology suggests that current approaches primarily focus on understanding user behavior, needs, and experiences, rather than deeply integrating cognitive or social mechanisms into RE models. Notably, some studies incorporate multiple disciplines. For instance, in \citePS{PS7}, both user experience and psychology are involved in the RE process. Similarly, \citePS{PS39} integrates HCI and user-centered design. This combination of disciplines indicates an emerging trend toward hybrid approaches, where multiple perspectives are combined to better capture complex human factors. However, such integrations remain limited and are not yet systematically structured.


\begin{table}[!h]
\small
\begin{tabular}{l l l l }
\hline
\textbf{Discipline}   & \textbf{PSs}                                                                                             & \textbf{\#} & \textbf{\%} \\ \hline
User centered design  & \citePS{PS3,PS10,PS15,PS20,PS25,PS27,PS31,PS32,PS35,PS39} & 10          & 25.64       \\ \hline
Psychology            & \citePS{PS7,PS13,PS26,PS45,PS50,PS51,PS56}                                          & 7           & 17.95       \\ \hline
Design thinking       & \citePS{PS1,PS4,PS9,PS19,PS55}                                                                  & 5           & 12.82       \\ \hline
HCI                   & \citePS{PS21,PS28,PS39,PS48}                                                               & 4           & 10.26       \\ \hline
Cognitive science    & \citePS{PS2,PS6,PS40}                                                                            & 3           & 7.69        \\ \hline
Interaction design    & \citePS{PS14,PS41,PS47}                                                                          & 3           & 7.69        \\ \hline
User experience       & \citePS{PS7,PS22}                                                                                      & 2           & 5.13        \\ \hline
Virtual reality       & \citePS{PS11,PS44}                                                                                     & 2           & 5.13        \\ \hline
Marketing Engineering & \citePS{PS29}                                                                                                & 1           & 2.56        \\ \hline
Social engineering    & \citePS{PS38}                                                                                                & 1           & 2.56        \\ \hline
Large Language Models    & \citePS{PS54}                                                                                                & 1           & 2.56        \\ \hline
\textbf{Total}                 &                                                                                                          & \textbf{39}          & \textbf{100.0 }      \\ \hline
\end{tabular}
\caption{Contributing disciplines involved in the RE process of PSs}
\label{tab:texternal-descipline}
\end{table}

\begin{figure}[]
  \centering
  \includegraphics[width=0.7\linewidth]{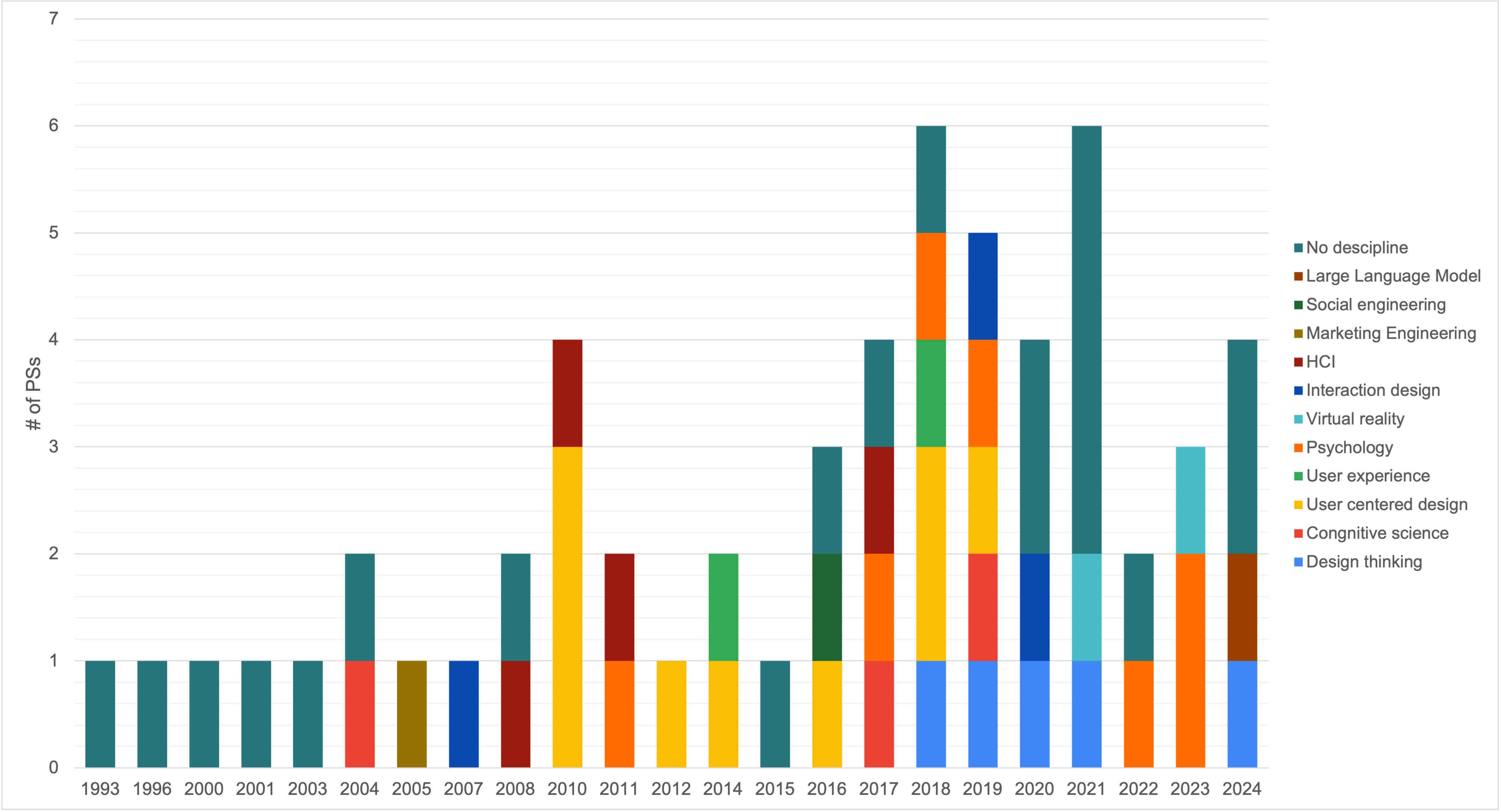}
  \caption{Temporal analysis of the Evolution of using contributing discipline in RE process }
  \label{fig:desciplineParAnnee}
\end{figure}

We conducted a temporal analysis to understand the emergence of using these disciplines in RE, as shown in Figure~\ref{fig:desciplineParAnnee}. Our analysis reveals that the first study involving contributing discipline was \citePS{PS2} in 2004, where cognitive science provided the framework for modeling users' intentions, beliefs, and desires. In 2005, the authors of \citePS{PS29} applied techniques from marketing engineering to identify user preferences across multiple product attributes, such as functionality, design, and price. By 2007, \citePS{PS47} adopted the persona technique from interaction design to pinpoint requirements specific to various user categories. From 2010 onwards, we observe an increased integration of contributing disciplines in requirements engineering. Notably, the first papers involving HCI \citePS{PS39,PS48} and user-centered design \citePS{PS10,PS32,PS40} were published in 2010. This trend continued with psychology \citePS{PS50} in 2011, user experience \citePS{PS22} in 2016, social engineering \citePS{PS38} in 2016, design thinking \citePS{PS4} in 2018, virtual reality \citePS{PS44} in 2021 and large language models~\citePS{PS54}. This temporal evolution shows a progressive expansion from cognitive and behavioral disciplines toward more interactive and technology-driven domains such as virtual reality and large language models. This shift suggests a transition from understanding user needs to actively shaping and simulating user experiences within the RE process.

\begin{tcolorbox}[colback=gray!10, colframe=black, title=Lessons Learned from RQ2]
\begin{itemize}
  \item Human-centered RE relies heavily on external disciplines, confirming its multidisciplinary nature.
  \item However, the integration remains dominated by design-oriented and behavioral disciplines, with limited incorporation of deeper cognitive and social theories.
  \item The recent emergence of technologies such as VR and LLMs indicates a shift toward more immersive and automated forms of user involvement.
\end{itemize}
\end{tcolorbox}

\subsection{RQ3: How do contributing disciplines complement the RE process?}

In this research question, we analyze the impact of each discipline on the RE process in terms of: 1) techniques used in the RE process (see table~\ref{tab:artefact}), 2) impact on the RE process (see table~\ref{tab:impact-on-RE-process}), and 3) transformation of the RE process to a human-centered RE process. This analysis aims to move beyond identifying disciplines to understanding how their contributions structurally transform the RE process.

\subsubsection{What techniques do contributing disciplines introduce?}



Table~\ref{tab:artefact} presents various techniques introduced into the Requirements Engineering (RE) process by contributing disciplines, categorized into three main goals:  \textbf{Understanding Human Cognition and Decision-Making}, \textbf{Enhancing User-Centered Design and Interaction}, and \textbf{Simulating and Automating Requirements Elicitation}. These categories reveal that contributing disciplines do not only introduce isolated techniques, but fundamentally extend RE along three complementary dimensions: understanding users, structuring interactions, and augmenting elicitation through technology.

Understanding Human Cognition and Decision-Making category involves disciplines that help in analyzing user behavior, mental processes, and decision-making models. For instance, Cognitive Science contributes through cognitive and mental models, while Psychology introduces psychological models, theories, and taxonomies to enhance understanding of user behavior and decision-making. Social Engineering provides scenario-based decision analysis, and Marketing Engineering applies conjoint analysis to identify user preferences, reinforcing the cognitive understanding aspect of RE. 

Enhancing the User-Centered Design and Interaction category involves disciplines that provide methods for integrating user needs, usability, and human-centric interactions in the RE process. These disciplines define usability standards, refine interaction techniques, and integrate human-centered design methodologies. For instance, user Experience contributes through interview techniques and UXD patterns, to ensure usability-driven requirements gathering. 

Simulating and Automating Requirements Elicitation involves emerging disciplines that focus on AI-driven automation, immersive environments, and advanced tools for capturing requirements efficiently. for instance, VR environments facilitate the simulation of user interactions, providing immersive scenarios for requirements validation. LLMs enhance RE by generating AI-driven personas and user profiles, automating persona creation and requirements elicitation.

Among these techniques, personas are widely used across multiple disciplines, including HCI, Interaction Design, and User-Centered Design, as a means to bridge the gap between technical requirements and user needs. A persona is a detailed, fictional representation of target users that provides insights into their goals, behaviors, and challenges~\cite{karolita2023use}. The widespread use of personas across disciplines indicates their role as a central bridging artifact, translating abstract human characteristics into actionable requirements representations.

\begin{table}[!h]
\begin{tabular}{llll}
\hline
\textbf{Proposed Technique}    & \textbf{Discipline}                                                                                                                    & \textbf{PSs}                                    & \textbf{\#PSs} \\ \hline
\multicolumn{4}{l}{\textbf{Understanding Human Cognition and Decision-Making}} \\ \hline
Cognitive Models      & Cognitive Science  & \citePS{PS6,PS40}  & 2     \\ \hline
Mental Models         & Cognitive Science  & \citePS{PS2}  & 1     \\ \hline
Psychological Models  & Psychology         & \citePS{PS7,PS13,PS26,PS50}  & 4     \\ \hline
Psychological Theories & Psychology        & \citePS{PS45, PS56}  & 2     \\ \hline
Taxonomies           & Psychology         & \citePS{PS51}  & 1     \\ \hline
Scenarios            & Social Engineering  & \citePS{PS38}  & 1     \\ \hline
Conjoint Analysis    & Marketing Engineering  & \citePS{PS29}  & 1     \\ \hline
\multicolumn{4}{l}{\textbf{Enhancing User-Centered Design and Interaction}} \\ \hline
Persona                & \begin{tabular}[c]{@{}l@{}}HCI, Interaction Design, \\ User-Centered Design\end{tabular}  & \citePS{PS3,PS10,PS21,PS32,PS35,PS41,PS48} & 7     \\ \hline
Processes             & \begin{tabular}[c]{@{}l@{}}Design Thinking, Interaction Design, \\ User-Centered Design, HCI\end{tabular}  & \citePS{PS1,PS4,PS9,PS14,PS19,PS20}  & 6     \\ \hline
HCI Design Patterns  & HCI  & \citePS{PS28,PS39}  & 2     \\ \hline
Comics               & Design Thinking  & \citePS{PS55}  & 1     \\ \hline
User Objectives      & User-Centered Design  & \citePS{PS27}  & 1     \\ \hline
User Stories         & User-Centered Design  & \citePS{PS25}  & 1     \\ \hline
Guidelines           & User-Centered Design  & \citePS{PS15}  & 1     \\ \hline
UCD Model            & User-Centered Design  & \citePS{PS31}  & 1     \\ \hline
Interview Techniques & User Experience  & \citePS{PS22}  & 1     \\ \hline
UXD Patterns         & User Experience  & \citePS{PS7}  & 1     \\ \hline
\multicolumn{4}{l}{\textbf{Simulating and Automating Requirements Elicitation}} \\ \hline
VR Environments      & Virtual Reality  & \citePS{PS11,PS44}  & 2     \\ \hline
AI-Generated Personas  & Large Language Models  & \citePS{PS54}  & 1     \\ \hline                                                                                                         
\textbf{Total}       &   &   & \textbf{38}    \\ \hline
\end{tabular}
\caption{Techniques Contributed by Various Disciplines in the RE Process (Sorted by Goal-Based Categories)}
\label{tab:artefact}
\end{table}

\subsubsection{How does each discipline impact the RE process?}

We analyzed how different disciplines influence the RE process. We distinguished three categories of approaches as shown in the table ~\ref{tab:impact-on-RE-process}.

The first category of approaches modifies an existing RE process. Several disciplines influence existing RE processes by integrating their methodologies. For example, Design Thinking \citePS{PS1} adds its steps (empathize, define, ideate, prototype, test) to the Scrum framework to improve user stories. User-Centered Design \citePS{PS10} brings user-centric techniques into the Tropos framework, focusing on early user involvement, active participation in the design process, function allocation between user and system, user feedback incorporation, and iterative design with prototypes. Similarly, Interaction Design \citePS{PS14} introduces steps like contextual analysis, interface engineering, and prototype evaluation for requirements elicitation. HCI \citePS{PS28,PS39,PS48} uses scenario-based design, HCI design patterns, and persona definition activities within requirements activities. Cognitive Science \citePS{PS40} adds cognitive engineering phases, while Interaction Design \citePS{PS47} includes persona and requirements value analysis. 
The second category of approaches suggests entirely new RE processes. Design Thinking \citePS{PS4,PS9} proposes processes with phases like exploration, alpha prototyping, and user tests, blending design thinking with Scrum. Cognitive Science \citePS{PS6} offers a three-phase process (contextualization, analysis, modeling) incorporating cognitive modeling to understand professional tasks. User-Centered Design \citePS{PS20,PS31} presents new processes that combine goal-oriented RE techniques with user-centered design tools, including object-oriented system analysis, model-based task analysis, and prototyping.
The third category of approaches imports a new process from other disciplines. For instance, Design Thinking \citePS{PS19} adopts the double diamond process for RE, which includes stages like data gathering (interviews, stakeholder maps), data synthesis (concept maps, personas, empathy maps), development (bodystorming), and delivery (prototyping) (see table~\ref{tab:impact-on-RE-process}). These three forms of impact indicate increasing levels of transformation, ranging from incremental adaptation of existing RE processes to the introduction of entirely new human-centered workflows.

\begin{table}[]
\begin{tabular}{|l|l|l|l|}
\hline
\textbf{\begin{tabular}[c]{@{}l@{}}Impact on \\ RE process\end{tabular}}                          & \textbf{PS} & \textbf{Details on impact}                                                                                                     & \textbf{Discipline}                                                     \\ \hline
                                                                                                  & \citePS{PS1}          & \begin{tabular}[c]{@{}l@{}}Integrates design thinking into Scrum, following  five phases: 1) empathize,\\ 2) define, 3) ideate, 4) prototype, and 5) test to define user stories.\end{tabular}                     & \begin{tabular}[c]{@{}l@{}}Design\\  thinking\end{tabular}              \\ \cline{2-4} 
                                                                                                  & \citePS{PS10}         & \begin{tabular}[c]{@{}l@{}}Uses user-centric techniques in Tropos framework,  including: 1) early user \\ focus, 2) active user  involvement, 3) user-system function allocation, \\ 4) integration of user feedback, and 5) iterative  prototyping.\end{tabular}                      & \begin{tabular}[c]{@{}l@{}}User \\ centered \\ design\end{tabular}      \\ \cline{2-4} 
                                                                                                  & \citePS{PS14}         & \begin{tabular}[c]{@{}l@{}}Enhances requirements elicitation with the SPID framework, incorporating \\
                                                                                                  three steps:  1) contextual analysis, 2) interface engineering,  and 3) prototype \\ evaluation.\end{tabular}                           & \begin{tabular}[c]{@{}l@{}}Interaction \\ design\end{tabular}           \\ \cline{2-4} 
                                                                                                  & \citePS{PS28}         & \begin{tabular}[c]{@{}l@{}}Applies scenario-based design in iterative cycles with HCI techniques: \\1) scenarios,  2) storyboards, and 3) prototypes to refine requirements \\ and design.\end{tabular}              & HCI                                                                     \\ \cline{2-4} 
                                                                                                  & \citePS{PS39}         & \begin{tabular}[c]{@{}l@{}}Extends scenario-based design by incorporating  HCI design patterns into \\iterative requirement  elicitation and user feedback cycles.\end{tabular}                         & \begin{tabular}[c]{@{}l@{}}HCI, User \\ centered \\ design\end{tabular} \\ \cline{2-4} 
                                                                                                  & \citePS{PS40}         & \begin{tabular}[c]{@{}l@{}}Introduces cognitive engineering into requirements elicitation to enhance \\ understanding of user cognition.\end{tabular}                            & \begin{tabular}[c]{@{}l@{}}Cognitive \\ modelling\end{tabular}          \\ \cline{2-4} 
                                                                                                  & \citePS{PS47}         & \begin{tabular}[c]{@{}l@{}}Integrates personas into scenario-based design  using two steps: 1) persona \\ analysis and 2) requirement value analysis.\end{tabular}                                     & \begin{tabular}[c]{@{}l@{}}Interaction \\ design\end{tabular}           \\ \cline{2-4} 
\multirow{-8}{*}{\begin{tabular}[c]{@{}l@{}}Modifying an \\ existing \\ RE process\end{tabular}} & \citePS{PS48}         & \begin{tabular}[c]{@{}l@{}}Adapts persona definition into requirements activities: 1) elicitation, \\ 2) analysis, 3)  specification, and 4) validation.\end{tabular}                       & HCI                                                                     \\ \hline
                                                                                                  & \citePS{PS4}          & \begin{tabular}[c]{@{}l@{}}Defines a new three-phase process:  1) exploration (design thinking),\\ 2) alpha  prototyping, and 3) user testing and market launch \\ (guided by Scrum).\end{tabular}                                       & \begin{tabular}[c]{@{}l@{}}Design \\ thinking\end{tabular}              \\ \cline{2-4} 
                                                                                                  & \citePS{PS6}          & \begin{tabular}[c]{@{}l@{}}Develops a three-phase process: 1) Contextualization, 2) analysis, and \\ 3) modeling, integrating  cognitive modeling to analyze professionals' \\ task performance in healthcare.\end{tabular}                              & \begin{tabular}[c]{@{}l@{}}Cognitive \\ science\end{tabular}           \\ \cline{2-4} 
                                                                                                  & \citePS{PS9}          & \begin{tabular}[c]{@{}l@{}}Combines Scrum and design thinking into an  eight-phase process:\\ 1) reframing, 2) analysis  and synthesis, 3) ideation, 4) planning, 5) \\ prototyping, 6) replanning, 7) development,  and 8) evaluation.\end{tabular}                         & \begin{tabular}[c]{@{}l@{}}Design \\ thinking\end{tabular}              \\ \cline{2-4} 
                                                                                                  & \citePS{PS20}         & \begin{tabular}[c]{@{}l@{}}Defines a four-phase process: 1) exploration,  2) problem identification, 3) \\ envisioning,  4) evaluation, using goal-oriented RE  and user-centered\\ design tools.\end{tabular}                            & \begin{tabular}[c]{@{}l@{}}User \\ centered \\ design\end{tabular}      \\ \cline{2-4} 
\multirow{-5}{*}{\begin{tabular}[c]{@{}l@{}}Proposing \\ a new RE \\ process\end{tabular}}       & \citePS{PS31}         & \begin{tabular}[c]{@{}l@{}}Develops a three-phase process: 1) classical  object-oriented analysis,\\ 2) model-based task analysis, and 3) prototyping  via executable system\\ models.\end{tabular} & \begin{tabular}[c]{@{}l@{}}User \\ centered \\ design\end{tabular}      \\ \hline

\multirow{0}{*}{\begin{tabular}[c]{@{}l@{}}A process \\ from another \\ discipline\end{tabular}}   &
  \citePS{PS19}  &
  \begin{tabular}[c]{@{}l@{}}Adapts the double diamond design  thinking process for RE: 1) data gathering \\ (interviews, stakeholder maps), 2) data  synthesis (concept maps, personas, \\ empathy  maps), 3) development (bodystorming),  and 4) delivery (prototypes).\end{tabular} &
  \begin{tabular}[c]{@{}l@{}}Design \\ thinking\end{tabular}    \\ \cline{2-4}
 &
  \citePS{PS55}  &
  \begin{tabular}[c]{@{}l@{}}Workshop-based structured process  aligned with design thinking: 1) problem \\ understanding, 2) solution exploration,  and 3) prototyping and testing.\end{tabular} 
  & \begin{tabular}[c]{@{}l@{}}Design \\ thinking\end{tabular}     \\ \hline
\end{tabular}

\caption{Impact of disciplines on the RE process}
\label{tab:impact-on-RE-process}

\end{table}

\subsubsection{How does each discipline transform RE into a human-centered process?}

We examined how each discipline contributes to the RE process and transforms it into a human-centered approach. 
Detailed analysis of each discipline's contribution to the RE process --- including
the specific techniques introduced, integration mode, and transformative effect ---
is provided in Tables~\ref{tab:impact-cognitive} through~\ref{tab:impact-MESE}
of Appendix~\ref{annex:disciplines-impact}. Across these contributions, a pattern
of complementarity emerges rather than redundancy. Psychology and cognitive science
operate primarily at the \textit{individual level}, providing models and theories
that explain user emotions, cognitive constraints, and decision-making processes ---
addressing \textit{why} users behave and feel the way they do. Design thinking, UCD,
and interaction design operate at the \textit{process level}, providing structured
methods such as empathy mapping, iterative prototyping, and contextual analysis that
embed user participation into RE workflows --- addressing \textit{how} to involve users
effectively. UX, HCI, and interaction design bridge both levels by offering concrete
artifacts such as personas, design patterns, and journey maps that translate
individual-level insights into process-level inputs. VR and LLMs represent an
emerging \textit{tool level}, automating or simulating aspects of user involvement
that were previously manual. This layered complementarity suggests that effective
human-centered RE benefits from combining disciplines across levels rather than
adopting any single discipline in isolation. Only a handful of primary studies ---
notably \citePS{PS7,PS39} --- explicitly combine multiple disciplines, indicating
that this integrative potential remains largely unrealized in practice. This layered
structure is the direct empirical precursor to the two-pathway taxonomy developed in
Section~\ref{sec:cross-rq-synthesis}.

\begin{tcolorbox}[colback=gray!10, colframe=black, title=Lessons Learned from RQ3]
\begin{itemize}
    \item Human-centered RE is structured as a layered integration of disciplines operating at individual, process, artifact, and tool levels.
    \item Current approaches combine these levels only partially, limiting the full potential of multidisciplinary integration.
    \item Effective human-centered RE requires coordinated integration across these levels rather than isolated use of individual disciplines.
\end{itemize}

\end{tcolorbox}

\subsection{RQ4: Which RE methodologies are used in human-centered approaches without external disciplines?}

In this research question, we analyze the primary studies that rely solely on the RE field to propose human-centered approaches. We identified the techniques proposed by each approach and examined how these methodologies transform the RE process into a human-centered one. Our analysis highlights that these approaches contribute to human-centered RE by incorporating emotional, social, organizational, linguistic, and cognitive aspects, which enhance user engagement, improve communication, and ensure system alignment with human needs and behaviors.

\subsubsection{User-Centric Requirements and Behavior-Driven RE}
A significant portion of the analyzed studies focuses on integrating emotional aspects into the RE process (see table~\ref{tab:methodo-RE}). Several methodologies embed emotional goals directly into goal models. For example, studies \citePS{PS23,PS30,PS33,PS34,PS36,PS42,PS49} extend the goal framework by incorporating additional dimensions such as emotional, cognitive, and workload considerations into system requirements.
A subset of these approaches—\citePS{PS23,PS30,PS33,PS34,PS49}—specifically extend goal frameworks by incorporating emotion-related requirements. These studies introduce structured methods to capture and model. However, they do not propose a systematic approach to transform emotional requirements into functional and quality goals, except for \citePS{PS34}, which proposes decomposing emotional goals into sub-goals and matching them to system characteristics that fulfill them. These studies primarily rely on extensive interviews and direct user engagement to elicit human-centered requirements, compensating for the absence of external disciplines such as psychology or cognitive science that traditionally provide structured explanations for aspects like emotions, motivation, and decision-making.

\subsubsection{Social and Organizational Contextualization}
Beyond individual user behavior, other studies emphasize the need to consider broader social, organizational, and cultural contexts in the RE process. These methodologies recognize that requirements are shaped not only by personal preferences but also by collective behaviors, work environments, and societal norms. For instance, approaches that focus on human, social, and organizational factors \citePS{PS12} highlight the necessity of embedding organizational structures, social interactions, and workplace constraints into RE processes. The goal is to uncover hidden requirements, implicit constraints, and unarticulated needs that arise in complex work environments. Similarly, studies that examine cultural differences in RE \citePS{PS18} reveal that user requirements are often shaped by regional, linguistic, and societal influences. 

\subsubsection{Linguistic and Cognitive Workload Considerations}
One of the major challenges in RE is ensuring that users, stakeholders, and developers share a mutual understanding of system requirements. For instance, near-natural specification languages \citePS{PS24} aim to bridge the gap between technical teams and end-users by allowing requirements to be expressed in a format that is intuitive and comprehensible to non-experts. 

Beyond linguistic accessibility, studies focusing on information flows \citePS{PS36}, workload, and communication \citePS{PS42} provide essential insights into how human agents interact with systems and handle cognitive complexity. Workload analysis techniques, for instance, identify potential areas where human users might experience excessive cognitive demands.

\begin{table}[]
\begin{tabular}{lll}
\hline
\textbf{PS} &
  \begin{tabular}[c]{@{}l@{}}Techniques proposed\\ by the approach\end{tabular} &
  \begin{tabular}[c]{@{}l@{}}How this methodology transforms RE into a \\ human-centered process\end{tabular} \\ \hline
\multicolumn{3}{l}{\textbf{Social and Organizational Contextualization}} \\ \hline
\citePS{PS12} &
  \begin{tabular}[c]{@{}l@{}}Human, social, and organizational \\ factors of users\end{tabular} &
  \begin{tabular}[c]{@{}l@{}}The HSO process highlights user, social, and organizational \\factors to uncover hidden services or system constraints.\end{tabular} \\ \hline
\citePS{PS18} &
  Cultural differences in RE &
  \begin{tabular}[c]{@{}l@{}}The findings highlight cultural characteristics \\ unique to requirements engineering across different regions.\end{tabular} \\ \hline
\citePS{PS43} &
  User segmentation &
  \begin{tabular}[c]{@{}l@{}}Segmenting users based on goal-based criteria to \\ better understand their characteristics and optimize \\ system usability and accessibility.\end{tabular} \\ \hline
\multicolumn{3}{l}{\textbf{User-Centric Requirements and Behavior-Driven RE}} \\ \hline
\citePS{PS13} &
  Acceptance requirements &
  \begin{tabular}[c]{@{}l@{}}A model and a meta-model that capture acceptance \\ and gamification knowledge, facilitating a systematic \\ acceptance requirements analysis based on gamification.\end{tabular} \\ \hline
\citePS{PS23} &
  Emotional goals &
  \begin{tabular}[c]{@{}l@{}}Motivational goal models incorporate emotional goals \\ to enhance user engagement and satisfaction.\end{tabular} \\ \hline
\citePS{PS30} &
  Goal models with emotional goals &
  \begin{tabular}[c]{@{}l@{}}Integrates emotional considerations from the early \\ requirements phase, ensuring they are not overlooked \\ in later development stages.\end{tabular} \\ \hline
\citePS{PS33} &
  Emotional requirements &
  \begin{tabular}[c]{@{}l@{}}Emotional requirements elicitation gathers real-time\\ user feedback on prototypes via multimedia and\\ semi-structured interviews.\end{tabular} \\ \hline
\citePS{PS34} &
  \begin{tabular}[c]{@{}l@{}}Transforming emotional\\ goals into quality and \\functional goals\end{tabular} &
  \begin{tabular}[c]{@{}l@{}}Ensures traceability of emotional goals by linking \\ them to functional and quality goals, allowing their \\ impact to be considered in system design.\end{tabular} \\ \hline
\citePS{PS46} &
  \begin{tabular}[c]{@{}l@{}}User desires and intentions \\ extracted from usage history\end{tabular} &
  \begin{tabular}[c]{@{}l@{}}Extracts user behavioral and environmental context \\ data to capture implicit requirements that might \\ not emerge in traditional requirement workshops.\end{tabular} \\ \hline
\citePS{PS49} &
  Positive and negative emotions &
  \begin{tabular}[c]{@{}l@{}}Captures and models emotions, developing detailed \\ scenarios based on emotional responses to improve \\ user experience and satisfaction.\end{tabular} \\ \hline
\citePS{PS53} &
  Human concern annotations &
  \begin{tabular}[c]{@{}l@{}}Human concern annotations link model elements to stakeholder \\perspectives, desired outcomes, and potential\\ issues, ensuring a user-centered approach.\end{tabular} \\ \hline
\multicolumn{3}{l}{\textbf{Linguistic and Cognitive Workload Considerations}} \\ \hline
\citePS{PS24} &
  \begin{tabular}[c]{@{}l@{}}Near-natural specification language \\ for requirements specification\end{tabular} &
  \begin{tabular}[c]{@{}l@{}}Near-natural specification language facilitates requirement \\definition  in a format easily understood  by development teams.\end{tabular} \\ \hline
\citePS{PS36} &
  \begin{tabular}[c]{@{}l@{}}Information flows between agents\\ in the system and its environment\end{tabular} &
  \begin{tabular}[c]{@{}l@{}}Models discourse and communication between agents \\ to improve understanding of collaborative processes \\ and cognitive workload in human-centered systems.\end{tabular} \\ \hline
\citePS{PS37} &
  \begin{tabular}[c]{@{}l@{}}Concepts of explicit incompleteness \\ and unspecified relations\end{tabular} &
  \begin{tabular}[c]{@{}l@{}}A semi-formal notation providing concepts  to represent \\vagueness, incompleteness, and  contradictions inherent\\ in user requirements  within socio-technical systems.\end{tabular} \\ \hline
\citePS{PS42} &
  \begin{tabular}[c]{@{}l@{}}Workload in terms of \\ communication and task handling\end{tabular} &
  \begin{tabular}[c]{@{}l@{}}Analyzes workload and cognitive complexity \\ to prevent system-induced overload on human users, \\ ensuring a balanced operational scenario.\end{tabular} \\ \hline
\end{tabular}
\caption{Human-centered approaches in RE.}
\label{tab:methodo-RE}
\end{table}

Comparing the RE-internal approaches identified in RQ4 with the discipline-driven approaches from RQ3 reveals an important asymmetry. The RE-internal approaches excel at modeling emotional and social constructs within established RE formalisms, for example by extending goal models with emotional goals~\citePS{PS30,PS34} or incorporating cultural factors~\citePS{PS18}. However, they lack the theoretical grounding that external disciplines provide for understanding \textit{why} these human factors matter. For instance, psychology offers validated models of emotion such as the Theory of Constructed Emotion~\citePS{PS45} that pure RE extensions do not leverage. Conversely, discipline-driven approaches bring rich theoretical foundations but often struggle to integrate seamlessly with existing RE frameworks and artifacts. This gap between theoretical depth and practical integration represents one of the most significant challenges for the field and directly motivates the conceptual framework proposed in Section~\ref{sec:recommendations}.

\begin{tcolorbox}[colback=gray!10, colframe=black, title=Lessons learned from RQ4]
\begin{itemize}
    \item RE-internal approaches successfully integrate human aspects (emotional, social, cognitive) into existing RE models.
  \item However, they rely mainly on empirical elicitation and lack strong theoretical grounding.
  \item This creates a gap between practical integration and theoretical understanding, limiting the maturity of human-centered RE.
\end{itemize}
\end{tcolorbox}

\subsection{RQ5: How are human-centered RE approaches evaluated?}

In this research question, we examine: 1) the evaluation methods of human-centered RE approaches and 2) the outcomes of the PSs.

\subsubsection{What evaluation methods are used? }
Our analysis revealed that only 39\% (22 out of 56) of the approaches have been evaluated. This low evaluation rate indicates that the field remains largely exploratory, with many approaches proposed without sufficient empirical validation. This highlights a significant gap in the validation of human-centered RE methodologies, raising concerns about the effectiveness and reliability of the remaining approaches that lack rigorous evaluation. In the following, we delve into the various evaluation methods employed by the studies, providing insights into how effectiveness is measured in the field of human-centered RE.

The table~\ref{tab:evaluation-methods} presents the various evaluation methods used to assess the effectiveness of human-centered approaches in RE across $56$ primary studies. The most frequently employed method is the case study, utilized by 13 studies. A case study is conducted to investigate a single entity or phenomenon in its real-life
context, within a specific time space~\cite{wohlin2012experimentation}.
For example, \citePS{PS12} evaluated the HSO process applied to software products in government sectors, focusing on automating processes and financial settlements, while \citePS{PS34} combined an industry case study with a user study to assess a mobile learning application, analyzing emotional goals and usability through participant feedback. \citePS{PS18} conducted a case study on cultural differences in RE, highlighting 11 cultural characteristics unique to Papua New Guinea’s indigenous culture. \citePS{PS45} assessed the impact of gamification and acceptance requirements through a case study involving various stakeholder groups. These case studies typically evaluated the practicality, usability, and impact of the proposed methodologies on real-world projects. This focus on usability and user satisfaction indicates that evaluation is centered on immediate user perception rather than long-term impact on RE outcomes.

User studies used in \citePS{PS6,PS28,PS30,PS34,PS38,PS39,PS42}, observe and analyze how real users interact with a system or approach, focusing on usability, user satisfaction, and feedback~\cite{buse2011benefits}.  For instance, \citePS{PS6} involved interviews with professionals to evaluate models representing work activities and technological support needs, ensuring that the models accurately reflected user requirements. \citePS{PS39} conducted two cycles of evaluation after the requirements exploration-design phase to capture user attitudes, system performance, and satisfaction through task completion and feedback. \citePS{PS42} evaluated tools and methods with 11 research staff, including those with HCI and RE experience, to gather comprehensive feedback on usability and practical application. These user studies often highlighted specific aspects such as task completion efficiency, user satisfaction, emotional response, and the practicality of integrating user feedback into the design process. 

Field studies, used by 1 study \citePS{PS29}, collect data in natural settings to provide insights into how methodologies perform in real-life conditions. Just like case studies, field studies are observational methods~\cite{wohlin2012experimentation}. In \citePS{PS29}, authors employed a field study to evaluate the proposed methodology by applying it to mobile phone product lines with over 150 users, providing insights into the methodology's effectiveness and revealing practical lessons learned from real-world application over a year-long period. This field study offered a broad perspective on how the methodology performed in a natural uncontrolled environment. However, the very limited use of field studies and controlled empirical evaluations highlights a lack of robust validation methods in the field.

Empirical evaluations, also used by 1 study \citePS{PS3}, involving systematic observation and measurement through the Goal-Question-Metric (GQM) framework. This study evaluated the efficiency and feasibility of an algorithm by testing goal achievability and planning outcomes for modeled personas.

Notably, two studies (\citePS{PS30,PS34}) utilized multiple evaluation methods. \citePS{PS30} combined a user study and a case study to evaluate emergency alarm systems. The user study involved comparing different notations and collecting qualitative feedback, while the case study modeled emotional, functional, and quality goals based on ethnographic data collected from interviews with participants. Similarly, \citePS{PS34} combined an industry case study with a user study to evaluate a mobile learning application. The case study analyzed emotional goals and usability through participant feedback, while the user study involved tasks and qualitative questions to evaluate the effectiveness of different analysis techniques.

\begin{table}[]
\begin{tabular}{llll}
\hline
\textbf{Evaluation   method}  & \textbf{PSs}                                                                                                                                                                                                    & \textbf{\# of PSs} & \textbf{\% of PSs} \\ \hline
Case study           & \citePS{PS4},\citePS{PS9},\citePS{PS12},\citePS{PS14},\citePS{PS18},\textbf{\citePS{PS30}},\textbf{\citePS{PS34}},\citePS{PS36},\citePS{PS37},\citePS{PS43},\citePS{PS45},\citePS{PS46},\citePS{PS47}                                                                                                                                        & 13        & 54\%     \\ \hline
User study           & \citePS{PS6},\citePS{PS28},\textbf{\citePS{PS30}},\textbf{\citePS{PS34}},\citePS{PS38},\citePS{PS39},\citePS{PS42},\citePS{PS54},\citePS{PS56}                                                                                                                                                                     & 9         & 38\%     \\ \hline
Field study          & \citePS{PS29}                                                                                                                                                                                                    & 1         & 4\%      \\ \hline
Empirical evaluation & \citePS{PS3}                                                                                                                                                                                                     & 1         & 4\%     \\ \hline
                     & \multicolumn{1}{r}{Total}                                                                                                                                                                     & 22        & 100\%    \\ \hline
No evaluation        & \begin{tabular}[c]{@{}l@{}}\citePS{PS1,PS2,PS5,PS7,PS8,PS10,PS11,PS13,PS15,PS16,PS17,PS19,PS20,PS21}, \\ \citePS{PS22,PS23,PS24,PS25,PS26,PS27,PS31,PS32,PS33,PS35,PS40,PS41} \\ \citePS{PS44,PS48,PS49,PS50,PS51}\end{tabular} & 34        & 61\%     \\ \hline
\end{tabular}

\caption{Evaluation of human-centered RE approaches}
\label{tab:evaluation-methods}
\end{table}

In evaluating the impact of human-centered approaches in RE, we analyzed primary studies to compare these approaches with traditional methods that do not consider the human aspects. Out of the 56 PSs, only a subset of 6 studies evaluated the human-centered impact in a comparative manner.  This lack of comparative evaluation makes it difficult to demonstrate the added value of human-centered approaches over traditional RE practices. This subset constitutes only 11.76\% of the total number of PSs and 30\% of the evaluated PSs. The findings reveal a generally positive impact on user satisfaction, requirement completeness, and overall process improvement. For instance, \citePS{PS9} demonstrated that Design Thinking combined with Scrum achieved 100\% user satisfaction and over 95\% completeness in requirements. \citePS{PS12} quantified the positive impact of the HSO process, identifying 23\% new or changed requirements in one case and 17\% in another. \citePS{PS30} highlighted the importance of addressing users' emotional needs in system design, resulting in an improved user experience with the developed prototype. \citePS{PS34} showed that the approach effectively identified functional and non-functional goals for addressing emotional needs. \citePS{PS39} illustrated the successful user engagement and satisfaction through the use of HCI patterns and UCD methods in health informatics tools. \citePS{PS43} emphasized improvements in the RE process quality, enhancing requirement consistency, correctness, and completeness by understanding potential and target users.


\subsubsection{What are the outcomes of the primary studies?}
Table~\ref{tab:outcomes} categorizes the outcomes of various studies on human-centered approaches in RE, providing a comprehensive overview of the frameworks, methods, modeling techniques, guidelines, and insights derived from the research. The "Frameworks and Methods" category shows a range of integrated frameworks and processes designed to enhance RE. Notable entries include the Design Thinking and Scrum Integrated Framework (\citePS{PS1,PS9}), the Human, Social, and Organizational (HSO) Framework for eliciting requirements \citePS{PS12}, and the Emotion-Oriented Goals (EMOG) Framework \citePS{PS26}. These frameworks aim to systematically incorporate human-centered elements into the RE process, addressing social, emotional, and organizational aspects.
The "Modeling and Specification Techniques" category highlights methods that facilitate a more nuanced understanding and representation of user requirements. Examples include the Visual Notation for Safety-Critical Systems \citePS{PS2} and the Persona-Based Modeling Approach \citePS{PS3}, which emphasize the importance of accurately capturing user contexts and personas in RE.
In the "Guidelines" category, studies like \citePS{PS14} and \citePS{PS25} provide practical recommendations for enhancing RE processes. The guidelines for the effectiveness of the SPID process from interaction design \citePS{PS14} and the application of user stories in assistive technology projects \citePS{PS25} offer actionable insights for practitioners.
Lastly, the "Insights" category reveals valuable findings on cultural characteristics, stakeholder empathy, and learning through social engineering. Studies such as \citePS{PS18,PS19} emphasize the importance of understanding user diversity and the context in which systems are deployed.
Overall, the categorization underscores the approaches and tools available for making RE more human-centered, highlighting the importance of integrating emotional, social, and organizational factors into the development process. This structured synthesis provides a valuable reference for researchers and practitioners aiming to enhance the human-centricity of RE methodologies.

\begin{table}[]
\begin{tabular}{|l|l|l|}
\hline
\textbf{Category} &
  \textbf{Studies} &
  \textbf{Details} \\ \hline
\multirow{21}{*}{Frameworks and Methods} &
  \citePS{PS1,PS9} &
  Design Thinking and Scrum Integrated Framework \\ \cline{2-3} 
 &
  \citePS{PS12} &
  \begin{tabular}[c]{@{}l@{}}Human, Social, and Organisational (HSO) Framework \\ for eliciting requirements\end{tabular} \\ \cline{2-3} 
 &
  \citePS{PS13} &
  \begin{tabular}[c]{@{}l@{}}Systematic Acceptance Requirements Analysis Framework \\ based on Gamification (Agon Framework)\end{tabular} \\ \cline{2-3} 
 &
  \citePS{PS17} &
  Design Process for Decision-Oriented Systems \\ \cline{2-3} 
 &
  \citePS{PS21} &
  Context-Based Persona Story Metamodel \\ \cline{2-3} 
 &
  \citePS{PS24} &
  Function Module System (FMs) for User-Oriented Analysis \\ \cline{2-3} 
 &
  \citePS{PS26} &
  Emotion-Oriented Goals (EMOG) Framework \\ \cline{2-3} 
 &
  \citePS{PS29} &
  \begin{tabular}[c]{@{}l@{}}Hanako Method Combining Persona-Scenario-Based \\ Analysis and Conjoint Analysis\end{tabular} \\ \cline{2-3} 
 &
  \citePS{PS30} &
  Goal Models Involving Emotional Goals \\ \cline{2-3} 
 &
  \citePS{PS33} &
  Emotional Goal Systematic Analysis Technique (EG-SAT) \\ \cline{2-3} 
 &
  \citePS{PS37} &
  SeeMe Diagramming Technique for Socio-Technical Systems \\ \cline{2-3} 
 &
  \citePS{PS40} &
  \begin{tabular}[c]{@{}l@{}}A Virtual Environment to Support the Societal participation \\ Education of Low-literates (VESSEL)\end{tabular} \\ \cline{2-3} 
 &
  \citePS{PS42} &
  Method and Tools for Analyzing Workload \\ \cline{2-3} 
 &
  \citePS{PS43} &
  User Segmentation in RE \\ \cline{2-3} 
 &
  \citePS{PS46} &
  \begin{tabular}[c]{@{}l@{}}Automated Method to Identify User Intentions using \\ inSitu Framework\end{tabular} \\ \cline{2-3} 
 &
  \citePS{PS48} &
  Incorporating and Adapting HCI Personas Technique \\ \cline{2-3} 
 &
  \citePS{PS52} &
  UrbanMobility Co-Design Canvas framework (UMCDC) \\ \cline{2-3} 
 &
  \citePS{PS53} &
  human-centered methodology for designing Cyber-Physical Systems \\ \cline{2-3} 
 &
  \citePS{PS54} &
  A persona generation tool \\ \cline{2-3} 
 &
  \citePS{PS55} &
  A method for collaborative vision creation (Cartooneering) \\ \cline{2-3} 
 &
  \citePS{PS56} &
  A gamified requirements engineering (RE) environment \\ \hline
\multirow{4}{*}{\begin{tabular}[c]{@{}l@{}}Modeling and Specification \\ Techniques\end{tabular}} &
  \citePS{PS2} &
  Visual Notation for Safety-Critical Systems \\ \cline{2-3} 
 &
  \citePS{PS3} &
  Persona-Based Modeling Approach \\ \cline{2-3} 
 &
  \citePS{PS10} &
  Annotated BPM Providing Semantics to BPM Activities \\ \cline{2-3} 
 &
  \citePS{PS36} &
  \begin{tabular}[c]{@{}l@{}}Method for Modelling Requirements for Complex \\ Sociotechnical Systems\end{tabular} \\ \hline
\multirow{2}{*}{Guidelines} &
  \citePS{PS13} &
  \begin{tabular}[c]{@{}l@{}}Guidelines for Effectiveness of SPID Process \\ from Interaction Design\end{tabular} \\ \cline{2-3} 
 &
  \citePS{PS25} &
  \begin{tabular}[c]{@{}l@{}}Results of the Application of User Stories in \\ Assistive Technologies Projects\end{tabular} \\ \hline
\multirow{3}{*}{Insights} &
  \citePS{PS18} &
  Cultural Characteristics of PNG’s Indigenous Culture \\ \cline{2-3} 
 &
  \citePS{PS19} &
  Techniques for Stakeholder Empathy and Understanding \\ \cline{2-3} 
 &
  \citePS{PS38} &
  Learning about Social Engineering Acts Through Gameplay \\ \hline
\end{tabular}
\caption{The outcomes of the PSs and their categories}
\label{tab:outcomes}
\end{table}

The low evaluation rate (39\%) raises concerns about the maturity of the field. The majority of proposed human-centered RE approaches have not been subjected to empirical scrutiny. Overall, this suggests that human-centered RE lacks standardized evaluation frameworks, limiting both its scientific maturity and its adoption in industrial practice. Moreover, among the evaluated studies, there is a notable absence of comparative evaluations: only 6 studies (11.76\%) explicitly compare human-centered approaches against traditional RE methods. This makes it difficult to assess whether the additional effort of integrating multidisciplinary techniques yields measurable improvements over conventional practices. The dominance of case studies (54\% of evaluated work) over controlled experiments also limits the ability to draw causal conclusions about the effectiveness of specific techniques. These observations point to a clear need for the community to invest in controlled, comparative evaluations that measure concrete outcomes such as requirement completeness, stakeholder satisfaction, and defect rates attributable to human-centered RE practices. 

Beyond evaluation frequency, the absence of shared operationalizations 
of human-centered outcomes points to a deeper problem. No primary study 
defines measurable criteria for what it means for a system to satisfy 
an experience requirement, a requirement specifying constraints on 
users' cognitive, emotional, or social states during interaction. This 
absence is precisely what separates current HC-RE practice from the 
XCRE vision: without the ability to specify and evaluate experience 
requirements formally, human-centered insights remain implicit and 
unverifiable throughout the RE lifecycle.

\begin{tcolorbox}[colback=gray!10, colframe=black, title=Lessons Learned from RQ5]
\begin{itemize}
    \item Human-centered RE approaches are weakly evaluated, with most studies relying on case studies and user studies.
    \item The lack of comparative and standardized evaluation limits the ability to assess their effectiveness against traditional RE methods.
    \item Stronger empirical validation frameworks are needed to support the maturity and adoption of human-centered RE.

\end{itemize}

\end{tcolorbox}

\subsection{Cross-RQ Synthesis: Two Integration Pathways in Human-Centered RE}
\label{sec:cross-rq-synthesis}
 
A cross-reading of all
five RQs together reveals a structural pattern that no single RQ captures:
the 56 primary studies organize into two parallel integration traditions that are opposite in method and almost entirely disconnected in practice. We term these
the \textbf{Cognitive-Formal (C-F) pathway} and the \textbf{Participatory-Iterative
(P-I) pathway}, and identify a small set of \textbf{Bridge studies} that span both.
This taxonomy is not imposed on the data; it emerges from three classification signals
that are already present in the RQ1--RQ5 tables and that, taken together, produce a
clean and consistent separation.
 
\subsubsection{Classification Criteria}
 
Each primary study is assigned to a pathway based on three criteria:
 
\begin{itemize}
    \item \textbf{Discipline orientation.} C-F studies draw on psychology, cognitive
    science, marketing engineering, social engineering, or RE-internal formal modeling.
    P-I studies draw on user-centered design (UCD), design thinking, human-computer
    interaction (HCI), interaction design, user experience (UX), virtual reality (VR),
    or large language models (LLMs).
 
    \item \textbf{Framework type.} When a study extends an existing RE framework,
    C-F studies exclusively use goal-based frameworks (i*, GRL, Tropos, motivational
    goal models, agent-oriented goal models). P-I studies exclusively use
    scenario-based frameworks (Agile, Scrum, use cases). This distinction holds without
    exception across all 56 primary studies.
 
    \item \textbf{Artifact and integration mode.} C-F studies produce formal
    notations, cognitive models, emotional goal extensions, and taxonomies.
    P-I studies produce personas, process adaptations, design guidelines, and
    prototyping workflows.
\end{itemize}
 
A study is classified as \textbf{Bridge} when it explicitly combines disciplines,
frameworks, or artifact types from both pathways as co-equal contributions, not when
one pathway element appears incidentally. Two studies (PS5, PS16) lacked sufficient
detail in the extracted data for classification and are marked unclassified.
Table~\ref{tab:pathway-classification} (Appendix~\ref{annex:pathway-classification}) presents the complete classification of all 56
primary studies across the two pathways.

\subsubsection{Pathway Statistics and Structural Findings}
 
Table~\ref{tab:pathway-stats} summarizes the distribution of key characteristics
across the two pathways and the Bridge group. Four structural findings emerge directly
from this distribution.

\begin{table}[h]
\small
\caption{Summary statistics by integration pathway.
C-F includes PS50 (n=25); Bridge excludes PS50 (n=5).
Two studies (PS5, PS16) are unclassified and excluded.}
\label{tab:pathway-stats}
\begin{tabular}{p{4.2cm} p{1.8cm} p{1.8cm} p{1.8cm}}
\hline
\textbf{Metric}
  & \textbf{C-F} \newline \textbf{(n=25)}
  & \textbf{P-I} \newline \textbf{(n=24)}
  & \textbf{Br}  \newline \textbf{(n=5)} \\ \hline
Evaluated PSs              & 13 (52\%) & 7 (29\%)  & 2 (40\%) \\
Goal-based framework       & 10        & 0         & 3        \\
Scenario-based framework   & 0         & 7         & 1        \\
No framework               & 15        & 17        & 1        \\

\hline

\end{tabular}
\end{table}

\textbf{Finding 1 : Near-equal distribution across pathways.}
Among the 54 classifiable studies, excluding 2 unclassified studies
(\citePS{PS5, PS16}) and counting \citePS{PS50} under the C-F pathway on the basis of its
conceptual-formal integration mechanism, the corpus yields 25 C-F studies
and 24 P-I studies, with 5 Bridge studies spanning both.
This near-equal distribution reflects equal investment by the field in two structurally distinct
integration strategies that have nonetheless developed largely
independently of one another.

\textbf{Finding 2 : Framework type as a pathway predictor.} Every study using a
goal-based framework is C-F or Bridge; every study using a scenario-based framework
is P-I or Bridge. This means the two most established families
of RE frameworks, goal modeling and scenario-based development, have been
developed by different disciplinary traditions that do not communicate. A practitioner
working within Tropos or i* encounters psychology and cognitive science contributions;
one working within Scrum or Agile encounters design thinking and UCD. A practitioner working within Tropos or i* will find psychology and cognitive science contributions; one working within Scrum or Agile will find design thinking and UCD contributions, and the two communities do not cross-reference each other.
 
\textbf{Finding 3 : Evaluation rate reversal.} C-F studies are evaluated at nearly
twice the rate of P-I studies (54\% vs.\ 29\%). This is counterintuitive, formal
approaches are typically assumed to be harder to evaluate, but it is consistent
with the data. Formal notations and goal model extensions produce discrete, testable
contributions amenable to case studies and user studies. P-I studies, which more
frequently propose new end-to-end methodologies, produce
contributions that are harder to evaluate comparatively. 
 
\textbf{Finding 4 : Multi-phase coverage advantage of C-F.} C-F studies show
slightly broader RE phase coverage (11 of 25 cover multiple phases) than P-I studies
(8 of 24). This means that the elicitation concentration identified in RQ1 is not a
property of human-centered RE in general, but a property of the P-I pathway
specifically. C-F approaches, anchored in goal modeling, tend naturally to span
elicitation, analysis, and specification within a single formal framework.
 
\subsubsection{The Five Bridge Studies}
 
The Bridge studies \citePS{PS3, PS7, PS10, PS20, PS56} are the most
consequential finding of this cross-RQ synthesis precisely because they are so rare.
Each combines disciplines, frameworks, or artifact types from both pathways:
\citePS{PS3} and \citePS{PS10} and \citePS{PS20} apply UCD disciplines (P-I) within
goal-based frameworks (C-F); \citePS{PS7} combines psychology (C-F theory) with UX
patterns (P-I artifact);  and \citePS{PS56} combines motivation theories from psychology
(C-F) with a gamification process (P-I).
 
Of the five Bridge studies, only two --- \citePS{PS3} and \citePS{PS56} have been
empirically evaluated. None has achieved widespread adoption or citation within either
pathway's primary community. This suggests that cross-pathway integration is not only
rare but structurally unrewarded: the field's review and citation patterns do not
surface bridge work as the exemplary contribution it represents. The five Bridge studies
should be treated as existence proofs that cross-pathway integration is methodologically
feasible, and as the primary starting points for the research agenda proposed in
Section~\ref{sec:recommendations}.
 
\begin{tcolorbox}[colback=gray!10, colframe=black,
    title=Lessons Learned from Cross-RQ Synthesis]
\begin{itemize}
    \item Human-centered RE has evolved into two structurally parallel and empirically
    disconnected traditions, each comprising 24 of the 56 primary studies.
    \item Framework type is a near-perfect pathway predictor: all goal-based frameworks
    appear in the C-F pathway, all scenario-based frameworks in P-I, with zero
    crossover across 56 studies.
    \item The field's central challenge is not developing more approaches within
    either pathway, but creating systematic methodological bridges between them.
    \item The structural separation between pathways, combined with the 
absence of Layer~3 translation mechanisms and standardized evaluation 
criteria, defines the gap that Experience-Centered Requirements Engineering 
(XCRE) must address: making user experience a first-class, formally 
specified, and empirically verifiable concern in RE.
\end{itemize}
\end{tcolorbox}

\section{Research Gaps}
\label{sec:gaps}

The cross-pathway analysis presented in Section~\ref{sec:cross-rq-synthesis} provides a structural explanation for the research gaps identified across individual research questions. The observed 24/24 split between the Cognitive-Formal (C-F) and Participatory-Iterative (P-I) pathways reveals that these gaps are not incidental but reflect a fundamental pattern of specialization. Each pathway has developed distinct strengths while neglecting complementary aspects addressed by the other. The near-absence of bridge studies, with only 5 out of 56 studies attempting integration and only 2 providing empirical evaluation, indicates that these complementary strengths have not been systematically combined. Consequently, the research gaps identified in this section should be understood as manifestations of pathway separation rather than isolated deficiencies. Addressing them requires deliberate cross-pathway integration rather than incremental refinement within individual traditions. 

A first critical gap concerns the concentration of existing approaches in the elicitation phase. Nearly half of the studies (48\%) focus on elicitation, reflecting the central role of user interaction in capturing requirements, with user involvement reported in 61\% of primary studies. In contrast, subsequent phases such as analysis (22\%) and specification (29\%) receive significantly less attention. This imbalance is particularly problematic in the context of human-centered requirements, as elicitation activities often generate insights related to emotions, values, and social identities that require systematic analysis and formalization. While several studies address the modeling of such aspects, a comprehensive approach that ensures their transformation into functional and quality requirements across the full lifecycle remains largely absent. Existing work provides only partial solutions, indicating a fundamental gap in lifecycle integration.

This limitation is strongly pathway-dependent. The concentration in elicitation is primarily driven by Participatory-Iterative approaches, which rely on scenario-based frameworks and iterative design processes that emphasize early-stage activities. In contrast, Cognitive-Formal approaches, anchored in goal-based modeling, demonstrate broader lifecycle coverage, with 11 out of 25 studies spanning multiple phases compared to 8 out of 24 for P-I approaches. This observation suggests that lifecycle discontinuity is not an inherent limitation of human-centered requirements engineering, but rather a structural consequence of the current dominance of P-I approaches in the corpus. It further indicates that goal-oriented modeling may provide a critical mechanism for extending human-centered insights beyond elicitation.

A second major gap concerns domain coverage. The literature shows a strong concentration in healthcare and e-learning, which are traditional domains for human-centered approaches. In contrast, domains characterized by complex human--technology interaction, such as assistive technologies, augmented reality, cyber-physical systems, and smart cities, remain significantly underexplored. This imbalance limits the generalizability of existing findings and indicates a critical need to extend HC-RE approaches to emerging domains where human-centered considerations are particularly essential.

A third gap relates to the integration of multidisciplinary contributions into the requirements engineering process. While multiple external disciplines have introduced methods and techniques that influence how requirements are elicited, analyzed, and validated, their integration remains partial and inconsistent. The influence of design thinking on agile practices, particularly Scrum, illustrates the potential of such integration, enabling iterative and user-focused development processes. However, these advances have not yet been systematically incorporated into the broader RE lifecycle. This highlights a fundamental need to redefine the RE process to fully integrate multidisciplinary and human-centered methodologies.

The cross-pathway analysis further reveals a critical risk associated with the wholesale adoption of processes from external disciplines. Among the studies that import complete processes, such as design frameworks or immersive environments, the majority belong to the Participatory-Iterative pathway, and none provide comparative evaluation against established RE methods. This pattern represents a high-risk mode of contribution, characterized by significant process transformation without empirical validation. Addressing this gap requires rigorous evaluation of imported methods before their adoption in industrial contexts.

A fourth gap concerns the limited exploration and validation of alternative elicitation techniques. Traditional methods, particularly interviews, remain dominant, while emerging approaches such as gaming, co-creation, and comic-based techniques are only marginally explored. The lack of systematic comparative studies prevents a clear understanding of their effectiveness, applicability, and limitations. Similarly, emerging AI-based techniques, including those based on large language models, remain insufficiently evaluated. This gap highlights the need for controlled empirical studies that compare traditional and innovative techniques across domains and user populations.

A fifth gap concerns the absence of comprehensive and generalizable frameworks that integrate human-centered approaches across all phases of the requirements engineering lifecycle, with only five Bridge studies~\citePS{PS3,PS7,PS10,PS20,PS56} address multiple phases, and these are typically limited to case-specific implementations rather than reusable methodologies. This indicates a fundamental lack of holistic frameworks capable of supporting end-to-end human-centered requirements engineering.

Finally, the identified Bridge studies provide critical insights into the feasibility of cross-pathway integration. These studies represent the only instances in which formal modeling and participatory processes are combined within a single approach. However, their limited number, lack of systematic evaluation, and absence of generalization highlight a major missed opportunity. These studies should be treated as foundational reference points for future work, enabling the systematic development of integrated HC-RE approaches rather than isolated contributions within individual pathways.

\section{Recommendations and Research Agenda}
\label{sec:recommendations}

The cross-pathway analysis presented in Section~\ref{sec:cross-rq-synthesis}
highlights that the main challenge in human-centered requirements engineering is not
the lack of approaches within individual traditions, but the absence of validated
integration between them. The Cognitive-Formal (C-F) pathway provides strong
theoretical foundations, while the Participatory-Iterative (P-I) pathway has driven
significant process innovation. However, neither has produced a fully integrated,
end-to-end HC-RE approach, and existing studies rarely connect the two in a systematic
manner. 

To address this limitation, this section proposes an organizing framework that
identifies where integration gaps occur, and derives a structured research agenda
organized into four priority tiers. Together, these elements provide a foundation for
advancing \textit{Experience-Centered Requirements Engineering (XCRE)}, in which user
experience is explicitly operationalized as a first-class concern in requirements
specification.

\subsection{The HC-RE Organizing Framework}
\label{sec:hcre-framework}

\begin{figure}[]
  \centering
  \includegraphics[width=1\linewidth]{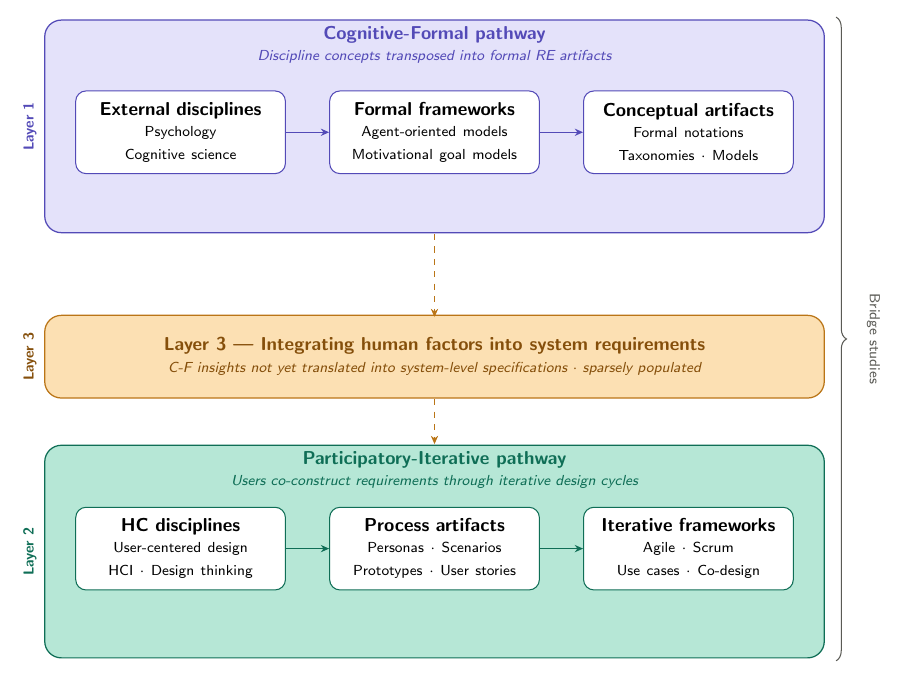}
  \caption{Organizing framework for the HC-RE research agenda. Layer~1 represents the
  theoretical core of the C-F pathway; Layer~2 captures the artifact and process core
  of the P-I pathway; Layer~3 corresponds to the integration layer, where insights
  must be translated into formal requirements.}
  \label{fig:hcse}
\end{figure}

The organizing framework illustrated in Figure~\ref{fig:hcse} synthesizes the
two-pathway taxonomy into a structured view of the field. Rather than prescribing a
specific methodology, it provides an explanatory model of how existing research is
distributed and where critical gaps remain. The framework is derived from the analysis
of 56 primary studies and reflects the layered structure underlying current HC-RE
approaches.

Layer~1 corresponds to the theoretical core of the C-F pathway. It draws on
disciplines such as psychology, cognitive science, marketing engineering, and social
engineering to explain \textit{why} users behave and feel in certain ways. Layer~2
represents the artifact and process core of the P-I pathway, relying on HCI, UX,
interaction design, design thinking, and UCD to define \textit{how} users are involved
in RE. 

Layer~3, however, remains largely underdeveloped. It represents the critical step of
translating insights from Layer~1 and artifacts from Layer~2 into formal RE
specifications, including functional and quality requirements. Only a small number of
studies explicitly address this translation. From an XCRE perspective, this layer is
essential, as it is where \textit{experience requirements}—that is, requirements
capturing constraints on users’ cognitive, emotional, or social states—must be
formalized and made verifiable.

To illustrate this gap, consider the following example. A C-F approach may identify
that users experience cognitive overload during a task (Layer~1), while a P-I approach
may represent this insight through personas or journey maps (Layer~2). What remains
missing is a systematic method to translate these insights into verifiable
requirements, such as constraints on interaction complexity or measurable cognitive
load thresholds. Addressing this gap is central to advancing HC-RE toward XCRE.

\subsection{Structured Research Agenda}
\label{sec:hcre-agenda}

Based on this framework, we identify four priority areas for future research. These
priorities reflect the key investments needed to bridge the C-F and P-I pathways and
to support the development of integrated, lifecycle-spanning HC-RE approaches.

\subsubsection{Priority 1: Cross-Pathway Integration Mechanisms}

The most pressing need is the development of mechanisms that connect the different
layers of the framework. This involves translating theoretical insights (e.g.,
cognitive models, emotional profiles) into design artifacts (e.g., personas, journey
maps), and subsequently into formal specifications. From an XCRE perspective, this
corresponds to transforming user experience into explicit, verifiable requirements.

Personas represent a promising bridge artifact, as they originate from participatory
approaches but encapsulate information that can be aligned with formal RE models.
Future work should aim to formalize personas as structured RE artifacts, with defined
schemas, traceability mechanisms, and integration rules linking them to goal models
and user stories. In addition, existing approaches that map emotional or experiential
concepts to functional and quality requirements should be generalized into reusable
methods applicable across different contexts.

\subsubsection{Priority 2: Lifecycle Integration Beyond Elicitation}

Current HC-RE approaches are strongly concentrated in the elicitation phase. This is
particularly evident in P-I approaches, which rely on scenario-based frameworks that
are less suited to later RE stages. In contrast, C-F approaches, often based on
goal-oriented models, tend to span multiple phases of the lifecycle.

Future research should therefore focus on extending human-centered practices beyond
elicitation. This includes ensuring that user-related insights—such as emotional
goals, values, and cognitive characteristics—are preserved, transformed, and validated
throughout analysis, specification, and validation. From an XCRE perspective, this
requires establishing traceability mechanisms that allow experience requirements to
be maintained and evaluated across the entire lifecycle.

\subsubsection{Priority 3: Empirical Maturation}

The field remains weakly validated from an empirical perspective. A limited proportion
of studies provide rigorous evaluation, and existing evaluation methods are often
inconsistent across approaches. Moreover, the C-F and P-I pathways rely on different
evaluation paradigms, making comparison difficult.

To address this issue, future work should develop shared evaluation frameworks with
clearly defined metrics, such as requirement completeness, stakeholder satisfaction,
and defect reduction. In the context of XCRE, these frameworks should also include
measures of user experience, such as perceived cognitive load, emotional response, and
usability. Particular attention should be given to emerging techniques and tools,
which require systematic validation in comparison with established methods.

\subsubsection{Priority 4: Domain and Technique Expansion}

Existing research is concentrated in a limited set of domains, particularly healthcare
and e-learning. Other domains characterized by complex human-technology interaction,
such as cyber-physical systems, assistive technologies, and smart environments, remain
underexplored.

Future research should extend HC-RE approaches to these domains by identifying their
specific human-centered requirements and adapting existing methods accordingly. In
parallel, a broader range of elicitation and interaction techniques should be explored
and evaluated in controlled settings. Comparative studies are needed to assess their
effectiveness with respect to different user populations, RE phases, and quality
criteria. Such efforts are essential to improve the generalizability and practical
applicability of human-centered RE approaches.

\section{Related Work}
\label{sec:related-work}

Table~\ref{tab:related-work-comparison} positions this study with respect to the most directly related secondary studies. The distinguishing dimension is the explicit focus on multidisciplinary integration. While prior work acknowledges the importance of human factors in requirements engineering (RE), no existing study systematically analyzes the contribution of external disciplines to RE as its primary unit of analysis.

\begin{table*}[!ht]
\footnotesize
\caption{Comparison of related secondary studies with this SMS.}
\label{tab:related-work-comparison}
\begin{tabular}{p{3.2cm} p{1.2cm} p{3.5cm} p{2.3cm} p{3.5cm}}
\hline
\textbf{Study} &
\textbf{Type} &
\textbf{Focus} &
\textbf{Multidisciplinary integration} &
\textbf{What is missing} \\ \hline

Hidellaarachchi et al.~\cite{hidellaarachchi2021effects} &
SLR &
Team-internal human aspects (personality, motivation, communication, culture, gender)
affecting RE practitioners; 74 studies across IEEE, ACM, Springer, Wiley &
Not addressed; explicitly identified as an open research direction &
Does not examine how external disciplines contribute methods to RE to capture
end-user needs; the gap our SMS directly addresses \\ \hline

Abelein \& Paech~\cite{abelein2015understanding} &
SMS &
Impact of user involvement on system outcomes &
Not addressed &
Does not explain which disciplines produce user involvement or why involvement
fails beyond elicitation \\ \hline

Wang et al.~\cite{wang2024uses} &
Empirical &
Persona adoption patterns in industrial RE &
Single technique only &
Does not analyze the disciplinary grounding of personas or their integration
across RE phases and framework types \\ \hline

Grundy et al.~\cite{grundy2021impact} &
Position paper &
Diversity of human characteristics in SE &
Identifies need; does not analyze it &
No systematic mapping; calls for integration without examining how it has occurred \\ \hline

\textbf{This study} &
\textbf{SMS} &
\textbf{How external disciplines integrate into RE to capture end-user needs,
emotions, and values} &
\textbf{Central focus: 11 disciplines, two-pathway taxonomy, 56 primary studies} &
\textbf{---} \\ \hline

\end{tabular}
\end{table*}

The systematic literature review of Hidellaarachchi et al.~\cite{hidellaarachchi2021effects} focuses on the individuals performing RE activities, analyzing how factors such as personality, motivation, communication, and organizational culture influence RE outcomes. In contrast, this study addresses a fundamentally different but complementary perspective by examining the disciplines that inform RE practices. Specifically, it analyzes how external fields contribute methods, models, and artifacts to capture end-user needs, emotions, and values. Together, these perspectives reflect the two dimensions of human factors in RE identified in Section~\ref{sec:background}.

Importantly, Hidellaarachchi et al.~explicitly identify cross-disciplinary collaboration as a critical and unresolved research direction. They highlight the need for RE researchers to extend beyond traditional software engineering knowledge and engage with other disciplines to improve the RE process~\cite{hidellaarachchi2021effects}. This study provides a direct and systematic response to that call by mapping, structuring, and analyzing how such cross-disciplinary contributions have been realized in practice.

Abelein and Paech~\cite{abelein2015understanding} demonstrate that user involvement has a positive impact on RE outcomes, but they also identify a fundamental limitation: the absence of methods that extend user involvement across multiple phases of the RE lifecycle. This limitation directly motivates the lifecycle analysis developed in RQ1 of this study. Wang et al.~\cite{wang2024uses} provide empirical insights into persona adoption in industrial contexts, identifying practical barriers but treating personas as an isolated technique. They do not analyze their theoretical foundations or their integration across RE phases and methodological frameworks. In contrast, this study situates personas within a broader multidisciplinary and lifecycle perspective through the proposed two-pathway taxonomy.

Grundy et al.~\cite{grundy2021impact} argue for the importance of integrating human characteristics into software engineering, but their contribution remains conceptual. They call for integration without providing a systematic analysis of how such integration has been achieved or operationalized. This limitation reflects a broader gap in the literature.

Taken together, prior work consistently confirms the importance of human factors in RE. However, it does not address three questions in a unified manner: which external disciplines contribute to RE, how their contributions differ in terms of integration mechanisms, and why these contributions remain fragmented. This study addresses these questions through a systematic mapping of 56 primary studies, the identification of 11 contributing disciplines, and the development of a two-pathway taxonomy that explains the structural separation between theoretical and process-oriented contributions.
\section{Threats to Validity}
\label{sec:threats-to-validity}

This section critically assesses the validity of the study following the framework proposed by Zhou et al.~\cite{zhou2016threats}. The objective is to explicitly identify potential limitations and to clarify the measures taken to mitigate their impact.

\subsection{Construct Validity}

Construct validity concerns the extent to which the search and data collection procedures accurately capture the intended body of knowledge. The primary search was conducted using Engineering Village (Compendex and Inspec), which index the major software engineering publishers, including IEEE, ACM, Springer, and Elsevier. This choice ensures strong coverage of high-quality and relevant venues.

The application of backward and forward snowballing identified nine additional studies, representing 16\% of the final corpus. The observed inclusion rate of 47\% suggests that the primary search achieved near-saturation, consistent with established snowballing practices~\cite{wohlin2014snowballing}. Nevertheless, a residual threat remains, as studies indexed exclusively in databases such as Scopus or Web of Science may not have been captured. Future replications should therefore extend the search strategy to include at least one additional database to further reduce this risk.

To mitigate interpretation bias, all cases in which primary studies did not explicitly report relevant information were systematically documented and cross-validated by at least two authors. This process reduces the risk of construct misinterpretation and ensures traceability of analytical decisions.

\subsection{Internal Validity}

Internal validity concerns the reliability and consistency of the study selection and data extraction processes. To ensure methodological rigor, the first and second authors independently conducted all three screening stages and performed data extraction for the full set of 56 primary studies. All disagreements were systematically resolved through discussion, and a third author was consulted when consensus could not be reached.

This full-corpus dual-review process, applied to every study rather than a subset, constitutes a strong control mechanism that minimizes selection and extraction bias. Although no formal inter-rater agreement statistic was computed, the exhaustive resolution of disagreements across the entire corpus is consistent with established guidelines for systematic mapping studies~\cite{petersen2015guidelines}.

The two-pathway taxonomy represents a post-hoc analytical construct and therefore introduces a potential classification bias. This threat was mitigated by defining explicit classification criteria and resolving all borderline cases collaboratively. The two studies that could not be reliably classified (PS5 and PS16) were excluded from all quantitative analyses, ensuring that reported results are not affected by uncertain categorizations.

\subsection{External Validity}

External validity concerns the generalizability of the findings. The corpus includes 56 studies published between 1993 and 2024, covering multiple domains, geographical contexts, and phases of the requirements engineering lifecycle. This diversity supports a broad level of representativeness and strengthens the generalizability of the identified patterns.

However, the restriction to English-language publications introduces a potential bias by excluding relevant contributions from non-English-speaking research communities. While this limitation is common in systematic studies, it remains a factor that may affect the completeness of the analysis.

\subsection{Conclusion Validity}

Conclusion validity concerns the robustness of the inferences drawn from the data. The study follows established guidelines for systematic mapping studies~\cite{petersen2015guidelines,keele2007guidelines}, and all methodological steps are fully documented in the publicly available replication package~\cite{benzarti2026replication}, supporting transparency and reproducibility.

The identification of the Layer~3 gap, based on a single primary study (\citePS{PS34}), represents a critical but preliminary finding. It should be interpreted as a well-supported hypothesis rather than a definitive conclusion. Targeted empirical validation is required to confirm its generality and to assess its implications for the design of integrated HC-RE methods.

\section{Discussion}

This study provides a structured and empirically grounded understanding of
human-centered requirements engineering (HC-RE) through the analysis of 56 primary
studies and the synthesis of their underlying patterns. The cross-RQ analysis shows
that the four dominant patterns identified in the literature—multidisciplinary
fragmentation, concentration on elicitation, the theory–practice gap, and limited
evaluation—are not independent issues. Rather, they reflect a deeper structural
condition: the separation between the Cognitive-Formal (C-F) and
Participatory-Iterative (P-I) pathways, which have evolved in parallel with little
systematic integration. 

Each pathway exhibits complementary strengths as well as important limitations.
The C-F pathway, based on goal-oriented frameworks and formal modeling, tends to
cover multiple phases of the RE lifecycle (11 out of 25 studies) and shows a higher
level of empirical evaluation (54\%). However, it provides limited support for
participatory processes. In contrast, the P-I pathway emphasizes iterative design
and user involvement, producing richer interaction processes and artifacts, but
remains largely concentrated in the elicitation phase and shows a lower evaluation
rate (29\%). The theory–practice gap identified in this study can be directly
explained by this separation: C-F approaches offer theoretical depth, while P-I
approaches provide practical integration mechanisms.

From the perspective of Experience-Centered Requirements Engineering (XCRE), this
structural separation has important implications. The C-F pathway models aspects of
user experience—such as cognition, emotion, and goals—but does not translate them
into verifiable system requirements. The P-I pathway captures user experience
through participatory artifacts, but does not formalize these insights within RE
models. As a result, user experience remains implicit and fragmented, and is rarely
preserved in specification and validation activities. Although a small number of
studies demonstrate the feasibility of cross-pathway integration, their limited
number and lack of strong empirical validation indicate that this direction remains
underdeveloped.

\subsection{Implications of the Two-Pathway Taxonomy}

The two-pathway taxonomy provides a concrete explanation of why user experience has
not yet become a first-class concern in requirements engineering. Several key
implications follow.

First, experience-related insights require explicit traceability mechanisms that
allow them to be carried from elicitation through to specification and validation.
Goal-oriented models provide a promising structural basis for such traceability,
and their combination with participatory practices represents a key direction for
future work.

Second, personas emerge as a central bridging artifact. While they originate in
participatory approaches, their content—user goals, values, and emotional
characteristics—can be aligned with formal RE representations. Formalizing personas
as structured and traceable artifacts would enable a more systematic transformation
of user insights into requirements.

Third, the lack of a shared evaluation framework across pathways limits the ability
to compare approaches and accumulate evidence. Establishing common evaluation
criteria—such as requirement completeness, user satisfaction, cognitive load, and
perceived usability—is essential for assessing the effectiveness of HC-RE methods.

Finally, existing cross-pathway studies should be treated as key reference points
for future research. Understanding how these studies combine theoretical and
participatory elements can inform the design of more integrated and scalable
approaches.

\subsection{Implications for Practice}

This study also provides several implications for practitioners.

First, human factors—including emotions, cognitive load, and user values—should not
only be elicited, but systematically incorporated into requirements analysis and
specification. This requires moving beyond early-stage user involvement and ensuring
that such insights are preserved throughout the development process.

Second, personas should be used as formal RE artifacts rather than informal design
tools. Structuring personas and linking them to goal models or user stories can
support traceability across development phases.

Third, practitioners should exercise caution when adopting approaches imported from
other disciplines, such as design thinking, virtual reality, or large language
models, without adequate empirical validation in RE contexts.

Finally, approaches grounded in goal-oriented modeling provide useful templates for
extending human-centered practices across the RE lifecycle, and can serve as a
foundation for more integrated methodologies.

\section{Conclusion}\label{sec:conclusion}

This paper presented a systematic mapping study of human-centered approaches in
requirements engineering, analyzing how these approaches are applied, which
disciplines contribute to them, and why their integration remains limited. The
analysis of 56 primary studies published between 1993 and 2024 shows that, although
the field increasingly draws on multidisciplinary knowledge, it remains structurally
fragmented across the RE lifecycle. 

The main contribution of this work is the identification of a two-pathway structure,
comprising Cognitive-Formal and Participatory-Iterative approaches. These two
traditions have developed in parallel, with limited interaction. The C-F pathway
offers theoretical depth and broader lifecycle coverage, while the P-I pathway
provides participatory processes and user-centered artifacts. However, neither has
produced a fully integrated HC-RE approach, Although five Bridge studies~\citePS{PS3,PS7,PS10,PS20,PS56} attempt
to bridge the gap between them.

Based on this analysis, the paper proposes a structured research agenda focusing on
four key priorities: cross-pathway integration mechanisms, lifecycle extension beyond
elicitation, empirical maturation, and domain expansion. These priorities define the
main steps required to advance toward Experience-Centered Requirements Engineering,
in which user experience is explicitly represented, operationalized, and validated
throughout the RE lifecycle.

Several research challenges remain. Human-centered approaches are still underexplored
in domains characterized by complex human–technology interactions, such as
assistive systems, augmented reality, and cyber-physical systems. Emerging
techniques—including gaming, co-creation, and AI-based persona generation—lack
systematic evaluation. Moreover, existing methods for translating human-centered
artifacts into formal requirements are limited and have not been widely replicated.

Addressing these challenges requires a shift from isolated methodological innovation
toward integrated, empirically validated approaches that connect disciplines,
artifacts, and RE phases. Such a shift is necessary to fully realize the potential of
human-centered requirements engineering and to establish user experience as a
first-class concern in software development.

\bibliographystylePS{elsarticle-num}
\bibliographyPS{references}

\bibliographystyle{elsarticle-num}
\bibliography{references}

\begin{thebibliography}{10}
\expandafter\ifx\csname url\endcsname\relax
  \def\url#1{\texttt{#1}}\fi
\expandafter\ifx\csname urlprefix\endcsname\relax\def\urlprefix{URL }\fi
\expandafter\ifx\csname href\endcsname\relax
  \def\href#1#2{#2} \def\path#1{#1}\fi

\bibitem{PS4}
J.~Hehn, F.~Uebernickel, The use of design thinking for requirements
  engineering: an ongoing case study in the field of innovative
  software-intensive systems, in: 2018 IEEE 26th international requirements
  engineering conference (RE), IEEE, 2018, pp. 400--405.

\bibitem{PS12}
A.~S. Andreou, Promoting software quality through a human, social and
  organisational requirements elicitation process, Requirements Engineering 8
  (2003) 85--101.

\bibitem{PS18}
P.~Yalamu, A.~Al~Mahmud, C.~Chua, Considerations for indigenous cultural
  aspects in software development: A case study, in: International Conference
  on Evaluation of Novel Approaches to Software Engineering, Springer, 2021,
  pp. 29--43.

\bibitem{PS25}
I.~C.~S. Zacharias, C.~Campese, T.~B. dos Santos, L.~P. da~Cunha, J.~M.~H.
  Costa, User stories method and assistive technology product development: A
  new approach to requirements elicitation, in: Proceedings of the Design
  Society: International Conference on Engineering Design, Vol.~1, Cambridge
  University Press, 2019, pp. 3781--3790.

\bibitem{PS31}
L.~Teixeira, C.~Ferreira, B.~S. Santos, User-centered requirements engineering
  in health information systems: A study in the hemophilia field, Computer
  methods and programs in biomedicine 106~(3) (2012) 160--174.

\bibitem{PS39}
A.~Sutcliffe, S.~Thew, O.~De~Bruijn, I.~Buchan, P.~Jarvis, J.~McNaught,
  R.~Procter, User engagement by user-centred design in e-health, Philosophical
  Transactions of the Royal Society A: Mathematical, Physical and Engineering
  Sciences 368~(1926) (2010) 4209--4224.

\bibitem{PS3}
G.~Nunes~Rodrigues, C.~Joel~Tavares, N.~Watanabe, C.~Alves, R.~Ali, A
  persona-based modelling for contextual requirements, in: Requirements
  Engineering: Foundation for Software Quality: 24th International Working
  Conference, REFSQ 2018, Utrecht, The Netherlands, March 19-22, 2018,
  Proceedings 24, Springer, 2018, pp. 352--368.

\bibitem{PS26}
S.~Alwidian, Towards extending the goal-oriented requirements language with
  emotion-oriented goals to support socio-technical systems, in: Proceedings of
  the 25th International Conference on Model Driven Engineering Languages and
  Systems: Companion Proceedings, 2022, pp. 306--311.

\bibitem{PS36}
A.~Sutcliffe, Requirements engineering for complex collaborative systems, in:
  Proceedings Fifth IEEE International Symposium on Requirements Engineering,
  IEEE, 2001, pp. 110--117.

\bibitem{PS42}
A.~Gregoriades, J.-E. Shih, A.~Sutcliffe, Human-centred requirements
  engineering, in: Proceedings. 12th IEEE International Requirements
  Engineering Conference, 2004., IEEE, 2004, pp. 154--163.

\bibitem{PS30}
T.~Miller, S.~Pedell, A.~A. Lopez-Lorca, A.~Mendoza, L.~Sterling, A.~Keirnan,
  Emotion-led modelling for people-oriented requirements engineering: the case
  study of emergency systems, Journal of Systems and Software 105 (2015)
  54--71.

\bibitem{PS34}
M.~Sherkat, T.~Miller, A.~Mendoza, R.~Burrows, Emotionalism within
  people-oriented software design, Frontiers in Computer Science 3 (2021)
  717787.

\bibitem{PS23}
E.~A. Oliveira, V.~Maram, L.~Sterling, Transitioning from motivational goal
  models to user stories within user-centred software design., in: RESOSY@
  APSEC, 2021.

\bibitem{PS33}
E.~Jackson, A.~Norta, Design of a remote emotional requirement elicitation
  feedback method, in: 2020 IEEE Third International Workshop on Affective
  Computing in Requirements Engineering (AffectRE), IEEE, 2020, pp. 3--8.

\bibitem{PS45}
T.~Iqbal, J.~G. Marshall, K.~Taveter, A.~Schmidt, Theory of constructed emotion
  meets re: An industrial case study, Journal of Systems and Software 197
  (2023) 111544.

\bibitem{PS49}
S.~F. Zulkifli, C.~W. Shiang, M.~A. bin Khairuddin, N.~bt~Jali, Modeling
  emotion oriented approach through agent-oriented approach, Int. J. Adv. Sci.
  Eng. Inf. Technol 10~(2) (2020) 647{\^a}.

\bibitem{PS20}
A.~Perini, N.~Seyff, M.~Stade, A.~Susi, Exploring re knowledge for
  gamification: Can re achieve a high score?, in: 2018 1st International
  Workshop on Affective Computing for Requirements Engineering (AffectRE),
  IEEE, 2018, pp. 14--19.

\bibitem{PS7}
P.~Kamthan, N.~Shahmir, Beyond utility and usability: towards affectability in
  agile software requirements engineering, in: 2018 International Conference on
  Computational Science and Computational Intelligence (CSCI), IEEE, 2018, pp.
  846--851.

\bibitem{PS21}
J.~Sede{\~n}o, E.-M. Sch{\"o}n, C.~Torrecilla-Salinas, J.~Thomaschewski, M.~J.
  Escalona, M.~Mejias, Modelling agile requirements using context-based persona
  stories, in: International Conference on Web Information Systems and
  Technologies, Vol.~2, SCITEPRESS, 2017, pp. 196--203.

\bibitem{PS27}
B.~Losada, Flexible requirement development through user objectives in an
  agile-ucd hybrid approach, in: Proceedings of the XIX International
  Conference on Human Computer Interaction, 2018, pp. 1--8.

\bibitem{PS1}
A.~Alhazmi, S.~Huang, Integrating design thinking into scrum framework in the
  context of requirements engineering management, in: Proceedings of the 3rd
  International Conference on Computer Science and Software Engineering, 2020,
  pp. 33--45.

\bibitem{PS9}
R.~d.~S. Braz, J.~R. Merlin, D.~Freitas Guilhermino~Trindade,
  C.~Eduardo~Ribeiro, E.~M. Sgarbi, F.~d.~S. Junior, Design thinking and scrum
  in software requirements elicitation: A case study, in: Design, User
  Experience, and Usability. Design Philosophy and Theory: 8th International
  Conference, DUXU 2019, Held as Part of the 21st HCI International Conference,
  HCII 2019, Orlando, FL, USA, July 26--31, 2019, Proceedings, Part I 21,
  Springer, 2019, pp. 179--194.

\bibitem{PS41}
A.~Dittmar, P.~Forbrig, Integrating personas and use case models, in:
  Human-Computer Interaction--INTERACT 2019: 17th IFIP TC 13 International
  Conference, Paphos, Cyprus, September 2--6, 2019, Proceedings, Part I 17,
  Springer, 2019, pp. 666--686.

\bibitem{PS47}
M.~Aoyama, Persona-scenario-goal methodology for user-centered requirements
  engineering, in: 15th IEEE International Requirements Engineering Conference
  (RE 2007), IEEE, 2007, pp. 185--194.

\bibitem{PS10}
C.~Di~Francescomarino, C.~Leonardi, A.~Marchetto, C.~D. Nguyen, N.~A. Qureshi,
  L.~Sabatucci, A.~Perini, A.~Susi, P.~Tonella, M.~Zancanaro, et~al., A bit of"
  persona", a bit of" goal", a bit of" process"... a recipe for analyzing user
  intensive software systems., in: iStar, 2010, pp. 36--40.

\bibitem{PS15}
H.~M.~T. Tran, F.~Anvari, A five-dimensional requirements elicitation framework
  for e-learning systems, International Journal of Information and Electronics
  Engineering 6~(3) (2016) 185.

\bibitem{PS32}
J.~D{\"o}rflinger, T.~Gross, Bottom billion architecture: An extensible
  software architecture for ict access in the rural developing world, in:
  Proceedings of the 4th ACM/IEEE International Conference on Information and
  Communication Technologies and Development, 2010, pp. 1--10.

\bibitem{PS35}
A.~A. Lopez-Lorca, T.~Miller, S.~Pedell, A.~Mendoza, A.~Keirnan, L.~Sterling,
  One size doesn't fit all: diversifying" the user" using personas and
  emotional scenarios, in: Proceedings of the 6th International Workshop on
  Social Software Engineering, 2014, pp. 25--32.

\bibitem{PS13}
L.~Piras, E.~Paja, P.~Giorgini, J.~Mylopoulos, Goal models for acceptance
  requirements analysis and gamification design, in: Conceptual Modeling: 36th
  International Conference, ER 2017, Valencia, Spain, November 6--9, 2017,
  Proceedings 36, Springer, 2017, pp. 223--230.

\bibitem{PS50}
R.~Proynova, B.~Paech, S.~H. Koch, A.~Wicht, T.~Wetter, Investigating the
  influence of personal values on requirements for health care information
  systems, in: Proceedings of the 3rd workshop on software engineering in
  health care, 2011, pp. 48--55.

\bibitem{PS51}
T.~Schweiß, L.~Thomaschewski, A.~Kluge, B.~Weyers, Software engineering for
  ar-systems considering user centered design approaches, Mensch und Computer
  2019 - Workshopband (2019).
\newblock \href {https://doi.org/10.18420/muc2019-ws-622}
  {\path{doi:10.18420/muc2019-ws-622}}.

\bibitem{PS56}
N.~Unkelos-Shpigel, B.~Berencwaig, S.~Kas, Revise that again: Are you
  motivated?, in: Proceedings of the 2nd International Workshop on Gamification
  in Software Development, Verification, and Validation, 2023, pp. 6--12.

\bibitem{PS19}
J.~C. de~Souza~Filho, W.~T. Nakamura, L.~M. Teixeira, R.~P. da~Silva, B.~F.
  Gadelha, T.~U. Conte, Towards a data-driven requirements elicitation tool
  through the lens of design thinking., in: ICEIS (2), 2021, pp. 283--290.

\bibitem{PS55}
S.~Storck, N.~Bartels, S.~A. Scherr, S.~Ludborzs, E.~Janke, Cartooneering-a
  collaborative workshop approach to develop scenarios with users., in: REFSQ
  Workshops, 2024.

\bibitem{PS28}
A.~Sutcliffe, S.~Thew, P.~Jarvis, Experience with user-centred requirements
  engineering, Requirements Engineering 16 (2011) 267--280.

\bibitem{PS48}
J.~W. Castro, S.~T. Acu{\~n}a, N.~Juristo, Integrating the personas technique
  into the requirements analysis activity, in: 2008 Mexican International
  Conference on Computer Science, IEEE, 2008, pp. 104--112.

\bibitem{PS2}
M.~Cebulla, Modeling concepts for safety-related requirements in sociotechnical
  systems, in: Computer Safety, Reliability, and Security: 23rd International
  Conference, SAFECOMP 2004, Potsdam, Germany, September 21-24, 2004.
  Proceedings 23, Springer, 2004, pp. 87--100.

\bibitem{PS6}
A.~Jatoba, A.~M. da~Cunha, M.~C. Vidal, C.~M. Burns, P.~V. de~Carvalho,
  Contributions from cognitive engineering to requirements specifications for
  complex sociotechnical systems: A case study in the context of healthcare in
  brazil, Human Factors and Ergonomics in Manufacturing \& Service Industries
  29~(1) (2019) 63--77.

\bibitem{PS40}
D.~G. Schouten, R.~T. Paulissen, M.~Hanekamp, A.~Groot, M.~A. Neerincx, A.~H.
  Cremers, Low-literates’ support needs for societal participation learning:
  Empirical grounding of theory-and model-based design, Cognitive Systems
  Research 45 (2017) 30--47.

\bibitem{PS14}
J.~C. Rosa, B.~B.~d. R{\^e}go, F.~A. Garrido, P.~D. Valente, N.~J. Nunes,
  E.~Matos, Interaction design and requirements elicitation integrated through
  spide: a feasibility study, in: Proceedings of the 19th Brazilian Symposium
  on Human Factors in Computing Systems, 2020, pp. 1--10.

\bibitem{PS22}
I.~Hastreiter, S.~Krause, T.~Schneidermeier, C.~Wolff, Developing ux for
  collaborative mobile prototyping, in: Design, User Experience, and Usability.
  Theories, Methods, and Tools for Designing the User Experience: Third
  International Conference, DUXU 2014, Held as Part of HCI International 2014,
  Heraklion, Crete, Greece, June 22-27, 2014, Proceedings, Part I 3, Springer,
  2014, pp. 104--114.

\bibitem{PS11}
N.~Chong, E.~Chu, A.~Nadonza, S.~M. Rodriguez, S.~Tith, J.~Shan, J.~Grundy,
  Y.~Wang, B.~Cheng, T.~Hoang, An empathetic approach to human-centric
  requirements engineering using virtual reality, in: 2023 IEEE 47th Annual
  Computers, Software, and Applications Conference (COMPSAC), IEEE, 2023, pp.
  1717--1724.

\bibitem{PS44}
Y.~Wang, B.~Cheng, T.~Hoang, C.~Arora, X.~Liu, Virtual reality enabled
  human-centric requirements engineering, in: 2021 36th IEEE/ACM International
  Conference on Automated Software Engineering Workshops (ASEW), IEEE, 2021,
  pp. 159--164.

\bibitem{PS29}
M.~Aoyama, Persona-and-scenario based requirements engineering for software
  embedded in digital consumer products, in: 13th IEEE International Conference
  on Requirements Engineering (RE'05), IEEE, 2005, pp. 85--94.

\bibitem{PS38}
K.~Beckers, S.~Pape, A serious game for eliciting social engineering security
  requirements, in: 2016 IEEE 24th International Requirements Engineering
  Conference (RE), IEEE, 2016, pp. 16--25.

\bibitem{PS54}
D.~Karolita, J.~Grundy, T.~Kanij, H.~O. Obie, J.~McIntosh, Crafter: A persona
  generation tool for requirements engineering., in: ENASE, 2024, pp. 674--683.

\bibitem{PS24}
M.~Li, A cooperative solving model supporting users-oriented requirements
  analysis, in: 1996 IEEE International Conference on Systems, Man and
  Cybernetics. Information Intelligence and Systems (Cat. No. 96CH35929),
  Vol.~2, IEEE, 1996, pp. 1568--1573.

\bibitem{PS43}
M.~Sherkat, T.~Miller, A.~Mendoza, Does it fit me better? user segmentation in
  requirements engineering, in: 2016 23rd Asia-Pacific Software Engineering
  Conference (APSEC), IEEE, 2016, pp. 65--72.

\bibitem{PS46}
J.~Yang, C.~K. Chang, H.~Ming, A situation-centric approach to identifying new
  user intentions using the mtl method, in: 2017 IEEE 41st Annual Computer
  Software and Applications Conference (COMPSAC), Vol.~1, IEEE, 2017, pp.
  347--356.

\bibitem{PS53}
T.~E. Jost, P.~Gr{\"u}nbacher, C.~Stary, Digital process twins for interleaving
  requirements elicitation and design of cyber-physical systems, in: 2024 IEEE
  32nd International Requirements Engineering Conference (RE), IEEE, 2024, pp.
  345--353.

\bibitem{PS37}
T.~Herrmann, M.~Hoffmann, K.-U. Loser, K.~Moysich, Semistructured models are
  surprisingly useful for user-centered design., in: COOP, 2000, pp. 159--174.

\bibitem{PS5}
I.~S. MacLeod, Scenario-based requirements capture for human factors
  integration, Cognition, Technology \& Work 10~(3) (2008) 191--198.

\bibitem{PS8}
L.~Costa, J.~Carneiro, M.~Temporao, Designing an app for nursing homes to
  clinical users, in: Proceedings of the 5th International Conference on
  Medical and Health Informatics, 2021, pp. 150--157.

\bibitem{PS16}
M.~Ehn, M.~Derneborg, {\AA}.~Reven{\"a}s, A.~Cicchetti, User-centered
  requirements engineering to manage the fuzzy front-end of open innovation in
  e-health: A study on support systems for seniors’ physical activity,
  International Journal of Medical Informatics 154 (2021) 104547.

\bibitem{PS17}
M.~L. Metersky, A decision-oriented approach to system design and development,
  IEEE transactions on systems, man, and cybernetics 23~(4) (1993) 1024--1037.

\bibitem{PS52}
Y.~Gebremichael, J.~Saad-Sulonen, A.~Knutas, Game on: Using serious tabletop
  games to enhance user engagement and optimize requirements engineering for
  smart city urban mobility solutions, in: Proceedings of the 2024
  International Conference on Information Technology for Social Good, 2024, pp.
  404--407.

\end{thebibliography}


\begin{thebibliography}{10}
\expandafter\ifx\csname url\endcsname\relax
  \def\url#1{\texttt{#1}}\fi
\expandafter\ifx\csname urlprefix\endcsname\relax\def\urlprefix{URL }\fi
\expandafter\ifx\csname href\endcsname\relax
  \def\href#1#2{#2} \def\path#1{#1}\fi

\bibitem{8559686}
IEEE, Iso/iec/ieee international standard - systems and software engineering --
  life cycle processes -- requirements engineering, ISO/IEC/IEEE 29148:2018(E)
  (2018) 1--104\href {https://doi.org/10.1109/IEEESTD.2018.8559686}
  {\path{doi:10.1109/IEEESTD.2018.8559686}}.

\bibitem{van2000requirements}
A.~Van~Lamsweerde, Requirements engineering in the year 00: A research
  perspective, in: Proceedings of the 22nd international conference on Software
  engineering, 2000, pp. 5--19.

\bibitem{nuseibeh2000requirements}
B.~Nuseibeh, S.~Easterbrook, Requirements engineering: a roadmap, in:
  Proceedings of the Conference on the Future of Software Engineering, 2000,
  pp. 35--46.

\bibitem{chakraborty2012role}
A.~Chakraborty, M.~K. Baowaly, A.~Arefin, A.~N. Bahar, The role of requirement
  engineering in software development life cycle, Journal of emerging trends in
  computing and information sciences 3~(5) (2012).

\bibitem{dalpiaz2013adaptive}
F.~Dalpiaz, P.~Giorgini, J.~Mylopoulos, Adaptive socio-technical systems: a
  requirements-based approach, Requirements engineering 18 (2013) 1--24.

\bibitem{snijders2015crowd}
R.~Snijders, A.~Ozum, S.~Brinkkemper, F.~Dalpiaz, Crowd-centric requirements
  engineering: A method based on crowdsourcing and gamification, Department of
  Information and Computing Sciences, Utrecht University, Tech. Rep. UU-CS-2015
  4 (2015).

\bibitem{boy2017human}
G.~A. Boy, Human-centered design of complex systems: An experience-based
  approach, Design Science 3 (2017) e8.

\bibitem{kasauli2021requirements}
R.~Kasauli, E.~Knauss, J.~Horkoff, G.~Liebel, F.~G. de~Oliveira~Neto,
  Requirements engineering challenges and practices in large-scale agile system
  development, Journal of Systems and Software 172 (2021) 110851.

\bibitem{hussain2020human}
W.~Hussain, H.~Perera, J.~Whittle, A.~Nurwidyantoro, R.~Hoda, R.~A. Shams,
  G.~Oliver, Human values in software engineering: Contrasting case studies of
  practice, IEEE Transactions on Software Engineering 48~(5) (2020) 1818--1833.

\bibitem{sutcliffe2022implications}
A.~Sutcliffe, P.~Sawyer, N.~Bencomo, The implications of
  ‘soft’requirements, in: 2022 IEEE 30th International Requirements
  Engineering Conference (RE), IEEE, 2022, pp. 178--188.

\bibitem{karolita2023use}
D.~Karolita, J.~McIntosh, T.~Kanij, J.~Grundy, H.~O. Obie, Use of personas in
  requirements engineering: A systematic mapping study, Information and
  Software Technology 162 (2023) 107264.

\bibitem{leonardi2011design}
C.~Leonardi, L.~Sabatucci, A.~Susi, M.~Zancanaro, Design as intercultural
  dialogue: coupling human-centered design with requirement engineering
  methods, in: Human-Computer Interaction--INTERACT 2011: 13th IFIP TC 13
  International Conference, Lisbon, Portugal, September 5-9, 2011, Proceedings,
  Part III 13, Springer, 2011, pp. 485--502.

\bibitem{sangiorgi2019human}
D.~Sangiorgi, F.~Lima, L.~Patr{\'\i}cio, M.~P. Joly, C.~Favini, A
  human-centred, multidisciplinary, and transformative approach to service
  science: a service design perspective, Handbook of Service Science, Volume II
  (2019) 147--181.

\bibitem{hidellaarachchi2021effects}
D.~Hidellaarachchi, J.~Grundy, R.~Hoda, K.~Madampe, The effects of human
  aspects on the requirements engineering process: A systematic literature
  review, IEEE Transactions on Software Engineering 48~(6) (2021) 2105--2127.
\newblock \href {https://doi.org/10.1109/TSE.2021.3055521}
  {\path{doi:10.1109/TSE.2021.3055521}}.

\bibitem{abelein2015understanding}
U.~Abelein, B.~Paech, Understanding the influence of user participation and
  involvement on system success--a systematic mapping study, Empirical Software
  Engineering 20~(1) (2015) 28--81.

\bibitem{wang2024uses}
Y.~Wang, C.~Arora, X.~Liu, T.~Hoang, V.~Malhotra, B.~Cheng, J.~Grundy, Who uses
  personas in requirements engineering: The practitioners' perspective, arXiv
  preprint arXiv:2403.15917 (2024).

\bibitem{grundy2021impact}
J.~C. Grundy, Impact of end user human aspects on software engineering., in:
  ENASE, 2021, pp. 9--20.

\bibitem{petersen2015guidelines}
K.~Petersen, S.~Vakkalanka, L.~Kuzniarz, Guidelines for conducting systematic
  mapping studies in software engineering: An update, Information and software
  technology 64 (2015) 1--18.

\bibitem{kitchenham2004procedures}
B.~Kitchenham, Procedures for performing systematic reviews, Keele, UK, Keele
  University 33~(2004) (2004) 1--26.

\bibitem{wohlin2012experimentation}
C.~Wohlin, P.~Runeson, M.~H{\"o}st, M.~C. Ohlsson, B.~Regnell, A.~Wessl{\'e}n,
  et~al., Experimentation in software engineering, Vol. 236, Springer, 2012.

\bibitem{keele2007guidelines}
S.~Keele, et~al., Guidelines for performing systematic literature reviews in
  software engineering, Tech. rep., Technical report, ver. 2.3 ebse technical
  report. ebse (2007).

\bibitem{darif2026cnl}
I.~Darif, G.~E. Boussaidi, S.~Kpodjedo, C.~Politowski, Controlled natural
  language for requirements specification: A systematic literature review, ACM
  Computing Surveys 58~(7) (2026) 1--36.
\newblock \href {https://doi.org/10.1145/3778169} {\path{doi:10.1145/3778169}}.

\bibitem{wohlin2014snowballing}
C.~Wohlin, Guidelines for snowballing in systematic literature studies and a
  replication in software engineering, in: Proceedings of the 18th
  International Conference on Evaluation and Assessment in Software
  Engineering, 2014, pp. 1--10.
\newblock \href {https://doi.org/10.1145/2601248.2601268}
  {\path{doi:10.1145/2601248.2601268}}.

\bibitem{benzarti2026replication}
I.~Benzarti, A.~Leshob, I.~Darif, H.~Mili, D.~Amayed, Two integration pathways
  in human-centered requirements engineering: A systematic mapping study of
  structural gaps, available at: \url{https://doi.org/10.5281/zenodo.20029273}
  (2026).
\newblock \href {https://doi.org/10.5281/zenodo.20029273}
  {\path{doi:10.5281/zenodo.20029273}}.

\bibitem{storey2020software}
M.-A. Storey, N.~A. Ernst, C.~Williams, E.~Kalliamvakou, The who, what, how of
  software engineering research: a socio-technical framework, Empirical
  Software Engineering 25 (2020) 4097--4129.

\bibitem{buse2011benefits}
R.~P. Buse, C.~Sadowski, W.~Weimer, Benefits and barriers of user evaluation in
  software engineering research, in: Proceedings of the 2011 ACM international
  conference on Object oriented programming systems languages and applications,
  2011, pp. 643--656.

\bibitem{zhou2016threats}
X.~Zhou, Y.~Jin, H.~Zhang, S.~Li, X.~Huang, A map of threats to validity of
  systematic literature reviews in software engineering, in: 2016 23rd
  Asia-Pacific Software Engineering Conference (APSEC), 2016, pp. 153--160.
\newblock \href {https://doi.org/10.1109/APSEC.2016.031}
  {\path{doi:10.1109/APSEC.2016.031}}.

\end{thebibliography}

\appendix

\section{Data extraction form fields}\label{annex:data-extraction}
The following table summarizes the eight thematic categories of the data extraction
form used to collect information from all 56 primary studies. The form comprised
27 sections and 52 questions in total; the complete list of individual questions
is available in the replication package~\cite{benzarti2026replication}.

\begin{table}[!h]
\small
\caption{Data extraction form: thematic categories and question counts}
\label{tab:extraction-categories}
\begin{tabular}{p{5.5cm} p{1.4cm} p{5.5cm}}
\hline
\textbf{Category} & \textbf{Questions} & \textbf{Key items covered} \\ \hline
General study information & 7 &
  Paper ID, year, title, authors, affiliation, source type, venue name \\ \hline
Research focus and framing & 5 &
  Paper goal, research questions, application domain, approach type, genericity \\ \hline
User and expert involvement & 4 &
  User involvement (Y/N), involvement techniques, expert involvement, type \\ \hline
RE phases and lifecycle scope & 2 &
  Targeted RE phase(s), RE-specific vs.\ full software lifecycle \\ \hline
Contributing disciplines and their impact & 7 &
  Disciplines involved (Y/N), impact on RE, techniques introduced, artifact types,
  transformation mechanism \\ \hline
RE framework integration & 4 &
  Framework extension (Y/N), model extended, framework type (goal-based /
  scenario-based) \\ \hline
RE-internal methodologies and techniques & 6 &
  Technique or methodology used, impact on RE, information type added,
  transformation mechanism, framework extension \\ \hline
Evaluation and outcomes & 17 &
  Evaluated (Y/N), evaluation method, human-centered impact assessed,
  impact direction, study outcomes, framework produced, citation count,
  search source (screening / snowballing) \\ \hline
\end{tabular}
\end{table}
\section{Detailed impact of the various disciplines}\label{annex:disciplines-impact}
The following tables provide a detailed breakdown of how each contributing discipline
transforms the RE process into a human-centered one, specifying for each primary study
the technique introduced, the type of information added, and the mechanism through
which the RE process is made more human-centered. These tables support the analytical
synthesis presented in RQ3 (Section~\ref{sec:findings}) and are referenced throughout
that section.
 
\begin{table}[H]
\begin{tabular}{llll}
\hline
\textbf{Technique} &
  \textbf{PS} &
  \textbf{\begin{tabular}[c]{@{}l@{}}What kind of information do these \\ disciplines add (model/tools/process)?\end{tabular}} &
  \textbf{\begin{tabular}[c]{@{}l@{}}How do these disciplines transform RE into a \\ human-centered process?\end{tabular}} \\ \hline
\begin{tabular}[c]{@{}l@{}}Situated \\ Cognitive\\ Engineering\end{tabular} &
  \citePS{PS40} &
  \begin{tabular}[c]{@{}l@{}}Iterative method for software \\ design and development.\end{tabular} &
  \begin{tabular}[c]{@{}l@{}}Captures context in societal \\ participation learning and translates \\ it into system specifications.\end{tabular} \\ \hline
\begin{tabular}[c]{@{}l@{}}Mental \\ model\end{tabular} &
  \citePS{PS2} &
  \begin{tabular}[c]{@{}l@{}}Represents user intentions, \\ beliefs, and desires.\end{tabular} &
  \begin{tabular}[c]{@{}l@{}}Enhances understanding of \\ human behavior in sociotechnical \\ systems.\end{tabular} \\ \hline
\begin{tabular}[c]{@{}l@{}}Cognitive \\ models\end{tabular} &
  \citePS{PS6} &
  \begin{tabular}[c]{@{}l@{}}Represents human capabilities \\ and task performance.\end{tabular} &
  \begin{tabular}[c]{@{}l@{}}Analyzes how healthcare \\ professionals perform \\ complex tasks.\end{tabular} \\ \hline
\end{tabular}
\caption{Impact of cognitive science on the human-centered RE process}
\label{tab:impact-cognitive}
\end{table}

\begin{table}[H]
\begin{tabular}{llll}
\hline
\textbf{Technique} &
  \textbf{PS} &
  \textbf{\begin{tabular}[c]{@{}l@{}}What is the kind of information these \\ disciplines add (model/tools/process)?\end{tabular}} &
  \textbf{\begin{tabular}[c]{@{}l@{}}How these disciplines transform RE to human \\ centric process?\end{tabular}} \\ \hline
\begin{tabular}[c]{@{}l@{}}Design \\ thinking \\ process\end{tabular} &
  \citePS{PS1} &
  Design thinking process to produce Epics &
  \begin{tabular}[c]{@{}l@{}}Integrated into Scrum to produce epics. If a \\ "Change Request" or "Problem request" occurs \\ during a sprint, the team decides whether to \\ start a new sprint or return to design thinking.\end{tabular} \\ \hline
\begin{tabular}[c]{@{}l@{}}Design \\ thinking \\ process\end{tabular} &
   \citePS{PS4} &
  \begin{tabular}[c]{@{}l@{}}Design thinking process to elicit needs\\  as mock-ups.\end{tabular} &
  \begin{tabular}[c]{@{}l@{}}Manages problem complexity through an \\ iterative approach. Elicitation techniques \\ (e.g., interviews, prototyping) are sequenced \\ to build a shared understanding and product vision.\end{tabular} \\ \hline
\begin{tabular}[c]{@{}l@{}}Design \\ thinking \\ process\end{tabular} &
  \citePS{PS9} &
  \begin{tabular}[c]{@{}l@{}}Unified process of Scrum and design \\ thinking\end{tabular} &
  Enhances Scrum with human-centered tasks. \\ \hline
\begin{tabular}[c]{@{}l@{}}Double \\ diamond \\ process\end{tabular} &
  \citePS{PS19} &
  \begin{tabular}[c]{@{}l@{}}Double Diamond process for RE: \\ 1) gather, 2) synthesize, 3) develop, \\ and 4) deliver.\end{tabular} &
  \begin{tabular}[c]{@{}l@{}}Uses empathy, collaboration, and \\ experimentation to align solutions with \\ user needs and drive innovation.\end{tabular} \\ \hline
Comics &
  \citePS{PS55} &
  \begin{tabular}[c]{@{}l@{}}Comics visualize user interactions with\\  the software.\end{tabular} &
  \begin{tabular}[c]{@{}l@{}}Engages users in shaping the software vision, \\ ensuring their needs and preferences \\ are reflected in the final product.\end{tabular} \\ \hline
\end{tabular}
\caption{Impact of design thinking on the human-centered RE process}
\label{tab:impact-design-thinking}

\end{table}

\begin{table}[H]
\begin{tabular}{llll}
\hline
\textbf{Technique}                                      & \textbf{PS} & \textbf{\begin{tabular}[c]{@{}l@{}}What kind of information do these \\ disciplines add (model/tools/process)?\end{tabular}}                                                                                                 & \textbf{\begin{tabular}[c]{@{}l@{}}How do these disciplines transform RE into a \\ human-centric process?\end{tabular}}                                                            \\ \hline
\begin{tabular}[c]{@{}l@{}}SPID\\ process\end{tabular} & \citePS{PS14}         & \begin{tabular}[c]{@{}l@{}}Sociocultural context of users and explores \\ problems and possible solutions.\end{tabular}                                                                                             & \begin{tabular}[c]{@{}l@{}}Involves users in requirements \\ elicitation.\end{tabular}                                                                 \\ \hline
Persona                                                & \citePS{PS41}         & \begin{tabular}[c]{@{}l@{}}Archetype of a user with detailed \\ needs, goals, and tasks.\end{tabular}                                                      & \begin{tabular}[c]{@{}l@{}}Fosters empathy and user \\ understanding for designers.\end{tabular} \\ \hline
Persona                                                & \citePS{PS47}         & \begin{tabular}[c]{@{}l@{}}Model of user classes with insights on \\ product usage and attitudes.\end{tabular} & \begin{tabular}[c]{@{}l@{}}Extracts requirements using \\ the Hanako method.\end{tabular}     \\ \hline
\end{tabular}
\caption{Impact of interaction design on the human-centered RE process}
\label{tab:impact-interaction-design}
\end{table}

\begin{table}[H]
\begin{tabular}{llll}
\hline
\textbf{Technique}                                           & \textbf{PS} & \textbf{\begin{tabular}[c]{@{}l@{}}What kind of information do these \\ disciplines add (model/tools/process)?\end{tabular}}                                                                           & \textbf{\begin{tabular}[c]{@{}l@{}}How do these disciplines transform RE \\ into a human-centric process?\end{tabular}}                                                                                                                                                                                                                   \\ \hline
Persona                                                     & \citePS{PS3}          & \begin{tabular}[c]{@{}l@{}}Personas defined by attributes, \\ goals, and contextual details.\end{tabular} & \begin{tabular}[c]{@{}l@{}}Gives users a face, making them \\ more tangible for design decisions.\end{tabular}                                                                                                                                           \\ \hline
Persona                                                     & \citePS{PS10}         & \begin{tabular}[c]{@{}l@{}}Persona descriptions include \\ user empathy and motivations.\end{tabular}                                                    & \begin{tabular}[c]{@{}l@{}}Using personas with scenarios enhances \\ system visualization and identifies \\ requirement issues.\end{tabular}                                                                              \\ \hline
\begin{tabular}[c]{@{}l@{}}Design\\ guidelines\end{tabular} & \citePS{PS15}         & Guidelines for designing questionnaires                                                                                                                                                                    & \begin{tabular}[c]{@{}l@{}}UCD methodology prioritizes \\ user goals.\end{tabular}                                                                                                                                                                                                                \\ \hline
\begin{tabular}[c]{@{}l@{}}User \\ stories\end{tabular}     & \citePS{PS25}         & \begin{tabular}[c]{@{}l@{}}User feedback on prototype \\ interactions.\end{tabular}                                                                                                              & \begin{tabular}[c]{@{}l@{}}Build prototypes based on user input \\ and refine user stories.\end{tabular}                                                                                                                 \\ \hline
\begin{tabular}[c]{@{}l@{}}TA\\ model\end{tabular}          & \citePS{PS31}         & \begin{tabular}[c]{@{}l@{}}Identifies sub-functionalities \\ within a task sequence.\end{tabular}                                & \begin{tabular}[c]{@{}l@{}}Involves users in understanding \\ system usability factors.\end{tabular} \\ \hline
Persona                                                     & \citePS{PS35}         & Persona's emotional scenario                                                                                                                                                                           & \begin{tabular}[c]{@{}l@{}}Uses scenarios to explore persona \\ reactions in different contexts.\end{tabular}                                                                                                                                              \\ \hline
\begin{tabular}[c]{@{}l@{}}User \\ objectives\end{tabular}  & \citePS{PS27}         & \begin{tabular}[c]{@{}l@{}}Defines functional and \\ non-functional user needs.\end{tabular}                                                                                         & \begin{tabular}[c]{@{}l@{}}Integrates agile and UCD \\ flexibly into projects.\end{tabular}                                                                                                                                                                 \\ \hline
\end{tabular}

\caption{Impact of UCD on the human-centered RE process}
\label{tab:impact-UCD}
\end{table}

\begin{table}[H]
\begin{tabular}{llll}
\hline
\textbf{Technique} &
  \textbf{PS} &
  \textbf{\begin{tabular}[c]{@{}l@{}}What is the kind of information these \\ disciplines add (model/tools/process)?\end{tabular}} &
  \textbf{\begin{tabular}[c]{@{}l@{}}How these disciplines transform RE into a \\ human-centric process?\end{tabular}} \\ \hline
\begin{tabular}[c]{@{}l@{}}Psychological\\ models\end{tabular} &
  \citePS{PS7} &
  \begin{tabular}[c]{@{}l@{}}Emotional aspects of requirements \\ modeling\end{tabular} &
  \begin{tabular}[c]{@{}l@{}}Ensures user experience by considering emotions \\ throughout development.\end{tabular} \\ \hline
\begin{tabular}[c]{@{}l@{}}Acceptance \\ meta-model\end{tabular} &
  \citePS{PS13} &
  \begin{tabular}[c]{@{}l@{}}Defines conditions for user system \\ acceptance.\end{tabular} &
  \begin{tabular}[c]{@{}l@{}}Enhances engagement by addressing psychological \\ factors affecting acceptance.\end{tabular} \\ \hline
\begin{tabular}[c]{@{}l@{}}Norman’s\\ model\end{tabular} &
  \citePS{PS26} &
  Emotion profiles of users &
  \begin{tabular}[c]{@{}l@{}}Recognizes variations in emotional needs \\ among stakeholders.\end{tabular} \\ \hline
\begin{tabular}[c]{@{}l@{}}Theory of\\ constructed \\ emotion\end{tabular} &
  \citePS{PS45} &
  \begin{tabular}[c]{@{}l@{}}Explains how emotions shape goals in \\ social settings.\end{tabular} &
  \begin{tabular}[c]{@{}l@{}}Supports design of emotive techniques for \\ stakeholder engagement.\end{tabular} \\ \hline
\begin{tabular}[c]{@{}l@{}}Psychological \\ models\end{tabular} &
  \citePS{PS50} &
  Personal values in user needs &
  \begin{tabular}[c]{@{}l@{}}Enhances persona and goal models \\ for deeper user understanding.\end{tabular} \\ \hline
\begin{tabular}[c]{@{}l@{}}Taxonomy \\ for AR\\ systems\end{tabular} &
  \citePS{PS51} &
  \begin{tabular}[c]{@{}l@{}}Four dimensions: social, technical, \\ teamwork, and benefits.\end{tabular} &
  \begin{tabular}[c]{@{}l@{}}Addresses new requirement types for \\ user-optimized AR systems.\end{tabular} \\ \hline
\begin{tabular}[c]{@{}l@{}}Motivation \\ theories\end{tabular} &
  \citePS{PS56} &
  \begin{tabular}[c]{@{}l@{}}Recognizes users as individuals with \\ distinct needs.\end{tabular} &
  \begin{tabular}[c]{@{}l@{}}Aligns RE with user motivation \\ to encourage participation.\end{tabular} \\ \hline
\end{tabular}

\caption{Impact of psychology on the human-centered RE process}
\label{tab:impact-psychology}
\end{table}

\begin{table}[H]
\begin{tabular}{llll}
\hline
\textbf{Technique} &
  \textbf{PS} &
  \textbf{\begin{tabular}[c]{@{}l@{}}What is the kind of information these\\ disciplines add (model/tools/process)?\end{tabular}} &
  \textbf{\begin{tabular}[c]{@{}l@{}}How these disciplines transform RE into a \\ human-centric process?\end{tabular}} \\ \hline
Persona &
  \citePS{PS21} &
  User needs and stakeholder values &
  \begin{tabular}[c]{@{}l@{}}Personas foster empathy for users and prevent \\ self-referential design.\end{tabular} \\ \hline
\begin{tabular}[c]{@{}l@{}}HCI \\ design \\ patterns\end{tabular} &
  \begin{tabular}[c]{@{}l@{}}\citePS{PS28}, \\ \citePS{PS39}\end{tabular} &
  \begin{tabular}[c]{@{}l@{}}Resolve design issues in early prototypes\\ and storyboards.\end{tabular} &
  \begin{tabular}[c]{@{}l@{}}Enable intuitive interfaces aligned with domain \\ models and user expectations.\end{tabular} \\ \hline
Persona &
  \citePS{PS48} &
  \begin{tabular}[c]{@{}l@{}}Activities to identify personas for \\ integration into SE.\end{tabular} &
  \begin{tabular}[c]{@{}l@{}}Personas model real user characteristics to \\ guide traditional RE activities.\end{tabular} \\ \hline
\end{tabular}

\caption{Impact of HCI on the human-centered RE process}
\label{tab:impact-hci}
\end{table}

\begin{table}[H]
\begin{tabular}{llll}
\hline
\textbf{Technique}                                      & \textbf{PS} & \textbf{\begin{tabular}[c]{@{}l@{}}What is the kind of information these \\ disciplines add (model/tools/process)?\end{tabular}}                                                                              & \textbf{\begin{tabular}[c]{@{}l@{}}How  these disciplines transform RE to human \\ centric process ?\end{tabular}}                                                            \\ \hline
SHIRA                                                  & \citePS{PS22}         & \begin{tabular}[c]{@{}l@{}}Abstract product qualities such as \\ 'controllable', 'simple',  'impressive' or \\ 'innovative' for a specific software \\ product in a specific context of use.\end{tabular} & \begin{tabular}[c]{@{}l@{}}Focus on UX methods to get a deep  insight not only \\ on pragmatic features but also emotional demands \\ (i.e. hedonic  qualities).\end{tabular} \\ \hline
\begin{tabular}[c]{@{}l@{}}UXD\\ patterns\end{tabular} & \citePS{PS7}          & \begin{tabular}[c]{@{}l@{}}UXD provides methodologies and \\ patterns to validate the prototypes \\ with users\end{tabular}                                                                               & \begin{tabular}[c]{@{}l@{}}UXD provides methodologies and patterns to \\ validate the prototypes with users\end{tabular}                                                      \\ \hline
\end{tabular}
\caption{Impact of UX on the human-centered RE process}
\label{tab:impact-UX}
\end{table}

\begin{table}[H]
\begin{tabular}{llll}
\hline
\textbf{Technique}                                        & \textbf{PS} & \textbf{\begin{tabular}[c]{@{}l@{}}What is the kind of information these \\ disciplines add (model/tools/process)?\end{tabular}} & \textbf{\begin{tabular}[c]{@{}l@{}}How  these disciplines transform RE to human \\ centric process ?\end{tabular}}                                                                                                                               \\ \hline
\begin{tabular}[c]{@{}l@{}}VR\\ environment\end{tabular} & \citePS{PS11}         & \begin{tabular}[c]{@{}l@{}}A simulation of the environment with \\ ADHD symptoms\end{tabular}                                 & \begin{tabular}[c]{@{}l@{}}VR can assist software engineers in better \\ understanding different viewpoints, challenges \\ and behaviours of diverse end users and stakeholders \\ through an empathetic approach.\end{tabular}                   \\ \hline
\begin{tabular}[c]{@{}l@{}}VR\\ environment\end{tabular} & \citePS{PS44}         & Simulation of persona                                                                                                        & \begin{tabular}[c]{@{}l@{}}VR can easily simulate different environments \\ and provide users with convenient interaction \\ with the software prototype and  immersive \\ experience so as to reveal human-centric \\ requirements.\end{tabular} \\ \hline
\end{tabular}
\caption{Impact of Virtual reality on the human-centered RE process}
\label{tab:impact-VR}
\end{table}

\begin{table}[H]
\begin{tabular}{lllll}
\hline
\textbf{Discipline} &
  \textbf{Technique} &
  \textbf{PS} &
  \textbf{\begin{tabular}[c]{@{}l@{}}What is the kind of information \\ these disciplines add (model/tools\\ /process)?\end{tabular}} &
  \textbf{\begin{tabular}[c]{@{}l@{}}How these disciplines transform \\ RE into a human-centric \\ process?\end{tabular}} \\ \hline
\begin{tabular}[c]{@{}l@{}}Marketing \\ engineering\end{tabular} &
  \begin{tabular}[c]{@{}l@{}}Conjoint \\ Analysis\end{tabular} &
  \citePS{PS29} &
  \begin{tabular}[c]{@{}l@{}}Identifies user preferences for \\ product attributes (e.g., function, \\ design, price).\end{tabular} &
  \begin{tabular}[c]{@{}l@{}}Segments users (personas) with \\ similar preferences.\end{tabular} \\ \hline
\begin{tabular}[c]{@{}l@{}}Social \\ engineering\end{tabular} &
  Scenarios &
  \citePS{PS38} &
  \begin{tabular}[c]{@{}l@{}}Models user behaviors and \\ game attack scenarios.\end{tabular} &
  \begin{tabular}[c]{@{}l@{}}Creates security training games that \\ simulate real-world social engineering \\ attacks and behavior patterns.\end{tabular} \\ \hline
\begin{tabular}[c]{@{}l@{}}Large \\ Language \\ Model\end{tabular} &
  Persona &
  \citePS{PS54} &
  \begin{tabular}[c]{@{}l@{}}Generates detailed persona \\ descriptions.\end{tabular} &
  \begin{tabular}[c]{@{}l@{}}Creates context-aware personas \\ by integrating human factors and \\ user-specific needs.\end{tabular} \\ \hline
\end{tabular}
\caption{Impact of Marketing engineering, social engineering and large language models on the human-centered RE process}
\label{tab:impact-MESE}
\end{table}

\section{Classification of All 56 Primary Studies by Integration Pathway}
\label{annex:pathway-classification}

The following table presents the complete two-pathway classification of all 56 primary
studies derived from the cross-RQ synthesis in Section~\ref{sec:cross-rq-synthesis}.
Studies are grouped by pathway --- Cognitive-Formal (C-F), Participatory-Iterative
(P-I), and Bridge --- and sorted within each group by primary discipline and then by
framework type, both alphabetically. The two unclassified studies (PS5, PS16) are
excluded from the table due to insufficient methodological detail in the extracted
data; their exclusion and its implications for the 24/24 symmetry claim are discussed
in Section~\ref{sec:cross-rq-synthesis}.

\begin{table*}[H]
\footnotesize
\caption{Classification of all 56 primary studies by integration pathway,
sorted by pathway and evaluation status.
C-F = Cognitive-Formal; P-I = Participatory-Iterative; Br = Bridge;
U = Unclassified.}
\label{tab:pathway-classification}
\begin{tabular}{p{0.7cm} p{0.6cm} p{5cm} p{4.8cm} p{1.4cm}}
\hline
\textbf{PS} & \textbf{Path} & \textbf{Primary Discipline(s)} &
\textbf{Framework Type} & \textbf{Evaluated} \\ \hline

\multicolumn{5}{l}{\textbf{Cognitive-Formal pathway}} \\ \hline
\citePS{PS6}  & C-F & Cognitive Science                       & None (new 3-phase process)   & Yes \\
\citePS{PS40} & C-F & Cognitive Science                       & None (situated cog.\ eng.)   & No  \\
\citePS{PS2}  & C-F & Cognitive Science                       & None (visual notation)       & No  \\
\citePS{PS29} & C-F & Marketing Engineering                   & None (conjoint analysis)     & Yes \\
\citePS{PS26} & C-F & Psychology                              & Goal (GRL)                   & No  \\
\citePS{PS45} & C-F & Psychology                              & Goal (POSE)                  & Yes \\
\citePS{PS13} & C-F & Psychology                              & Goal (Tropos)                & No  \\
\citePS{PS51} & C-F & Psychology                              & None (taxonomy)              & No  \\
\citePS{PS18} & C-F & RE-internal (cultural analysis)         & None                         & Yes \\
\citePS{PS17} & C-F & RE-internal (decision modeling)         & None                         & No  \\
\citePS{PS30} & C-F & RE-internal (emotional goals)           & Goal (agent-oriented)        & Yes \\
\citePS{PS23} & C-F & RE-internal (emotional goals)           & Goal (motivational)          & No  \\
\citePS{PS34} & C-F & RE-internal (emotional goals → formal)  & Goal (agent-oriented)        & Yes \\
\citePS{PS33} & C-F & RE-internal (emotional requirements)    & Goal (POSE)                  & No  \\
\citePS{PS53} & C-F & RE-internal (human concern annotations) & None                         & No  \\
\citePS{PS36} & C-F & RE-internal (information flows)         & Goal (i*)                    & Yes \\
\citePS{PS24} & C-F & RE-internal (linguistic specification)  & None                         & No  \\
\citePS{PS49} & C-F & RE-internal (people-oriented SE)        & Goal (Tropos)                & No  \\
\citePS{PS37} & C-F & RE-internal (semi-formal notation)      & None (SeeMe)                 & Yes \\
\citePS{PS12} & C-F & RE-internal (social/organizational)     & None (HSO process)           & Yes \\
\citePS{PS46} & C-F & RE-internal (usage history mining)      & None                         & Yes \\
\citePS{PS43} & C-F & RE-internal (user segmentation)         & None                         & Yes \\
\citePS{PS42} & C-F & RE-internal (workload/cognitive load)   & Goal (i*)                    & Yes \\
\citePS{PS38} & C-F & Social Engineering                      & None (game scenarios)        & Yes \\ 
\citePS{PS50} & C-F & Psychology (personal values) & None & No \\\hline

\multicolumn{5}{l}{\textbf{Participatory-Iterative pathway}} \\ \hline
\citePS{PS8}  & P-I & Design Thinking                         & None (full SE lifecycle)     & No  \\
\citePS{PS1}  & P-I & Design Thinking                         & Scenario (Scrum)             & No  \\
\citePS{PS4}  & P-I & Design Thinking                         & Scenario (Scrum)             & Yes \\
\citePS{PS9}  & P-I & Design Thinking                         & Scenario (Scrum)             & Yes \\
\citePS{PS55} & P-I & Design Thinking (comics)                & None                         & No  \\
\citePS{PS19} & P-I & Design Thinking (double diamond)        & None                         & No  \\
\citePS{PS48} & P-I & HCI                                     & None (persona into RE)       & No  \\
\citePS{PS21} & P-I & HCI                                     & None (persona metamodel)     & No  \\
\citePS{PS28} & P-I & HCI                                     & None (scenario-based design) & Yes \\
\citePS{PS39} & P-I & HCI + UCD                               & None (scenario-based)        & Yes \\
\citePS{PS14} & P-I & Interaction Design                      & None (SPID framework)        & Yes \\
\citePS{PS41} & P-I & Interaction Design                      & Scenario (use cases)         & No  \\
\citePS{PS47} & P-I & Interaction Design                      & Scenario (use cases)         & Yes \\
\citePS{PS54} & P-I & Large Language Models                   & None                         & Yes \\
\citePS{PS52} & P-I & RE-internal (co-design canvas)          & None                         & No  \\
\citePS{PS15} & P-I & UCD                                     & None (design guidelines)     & No  \\
\citePS{PS32} & P-I & UCD                                     & None (gaming, co-creation)   & No  \\
\citePS{PS31} & P-I & UCD                                     & None (new process)           & No  \\
\citePS{PS35} & P-I & UCD                                     & None (persona + scenarios)   & No  \\
\citePS{PS25} & P-I & UCD                                     & Scenario (user stories)      & No  \\
\citePS{PS27} & P-I & UCD                                     & Scenario (Agile)             & No  \\
\citePS{PS22} & P-I & UX                                      & None (SHIRA)                 & No  \\
\citePS{PS11} & P-I & Virtual Reality                         & None                         & No  \\
\citePS{PS44} & P-I & Virtual Reality                         & None (VR persona simulation) & No  \\ \hline

\multicolumn{5}{l}{\textbf{Bridge studies}} \\ \hline
\citePS{PS56} & Br  & Psychology + gamification process       & None                         & Yes \\
\citePS{PS7}  & Br  & Psychology + UX                         & Scenario (Agile)             & No  \\
\citePS{PS3}  & Br  & UCD + goal-based framework              & Goal (contextual)            & Yes \\
\citePS{PS10} & Br  & UCD + goal-based framework              & Goal (Tropos)                & No  \\
\citePS{PS20} & Br  & UCD + goal-based framework              & Goal (Tropos)                & No  \\ \hline

\multicolumn{5}{l}{\textbf{Unclassified}} \\ \hline
\citePS{PS5}  & U   & None identified                         & None                         & No  \\
\citePS{PS16} & U   & None identified                         & None                         & No  \\ \hline

\end{tabular}
\end{table*}

\printcredits
\section*{Declaration of Generative AI and AI-assisted technologies in the writing process}
During the preparation of this work, the authors used AI-assisted tools for language editing and formatting purposes. The authors reviewed and edited all content as needed and take full responsibility for the content of the publication.

\section*{Declaration of competing interest}
The authors declare that they have no known competing financial interests or personal relationships that could have appeared to influence the work reported in this paper.

\section*{Data availability}
The replication package including the search results, screening decisions, and data extraction forms is available at: \url{https://doi.org/10.5281/zenodo.20029273}~\cite{benzarti2026replication}.

\end{document}